\documentclass[fleqn,usenatbib]{mnras}

\usepackage{newtxtext,newtxmath}
\usepackage{multirow}
\usepackage{booktabs}
\usepackage[T1]{fontenc}

\DeclareRobustCommand{\VAN}[3]{#2}
\let\VANthebibliography\thebibliography
\def\thebibliography{\DeclareRobustCommand{\VAN}[3]{##3}\VANthebibliography}

\usepackage{graphicx}	% Including figure files
\usepackage{amsmath}	% Advanced maths commands
\title[A molecular ring in AG Car]{A warm molecular ring in AG Car: composing the mass-loss puzzle}

\author[C. Bordiu et al.]{C. Bordiu,$^{1,3}$\thanks{E-mail: cristobal.bordiu@inaf.it}
F. Bufano,$^{1}$
L. Cerrigone,$^{4}$
G. Umana,$^{1}$
J. R. Rizzo,$^{2,3}$
C. S. Buemi,$^{1}$
\newauthor
P. Leto,$^{1}$
F. Cavallaro,$^{1,5,6}$
A. Ingallinera,$^{1}$
S. Loru,$^{1}$
C. Trigilio,$^{1}$
S. Riggi$^{1}$
\\
$^{1}$INAF-Osservatorio Astrofisico di Catania, Via Santa Sofia 78, 95123 Catania\\
$^{2}$ISDEFE, Beatriz de Bobadilla 3, E-28040 Madrid, Spain\\
$^{3}$Centro de Astrobiolog\'ia (INTA-CSIC), Ctra. M-108, km. 4, E-28850 Torrej\'on de Ardoz, Madrid, Spain \\
$^{4}$Joint ALMA Observatory, Alonso de C\'ordova 3107, Vitacura, Santiago, Chile\\
$^{5}$Inter-University Institute for Data Intensive Astronomy, Cape Town, South Africa\\
$^{6}$Department of Astronomy, University of Cape Town, Cape Town, South Africa\\
}

\date{Accepted XXX. Received YYY; in original form ZZZ}

\pubyear{2020}

\begin{document}
\label{firstpage}
\pagerange{\pageref{firstpage}--\pageref{lastpage}}
\maketitle

% Abstract of the paper
\begin{abstract}
We present APEX observations of CO $J=3\rightarrow2$ and ALMA observations of CO $J=2\rightarrow1$, $^{13}$CO $J=2\rightarrow1$ and continuum toward the galactic luminous blue variable AG Car. These new observations reveal the presence of a ring-like molecular structure surrounding the star. Morphology and kinematics of the gas are consistent with a slowly expanding torus located near the equatorial plane of AG Car. Using non-LTE line modelling, we derived the physical parameters of the gas, which is warm ($\sim50$ K) and moderately dense ($\sim10^3$ cm$^{-3}$). The total mass of molecular gas in the ring is $2.7\pm0.9$ M$_{\sun}$. We analyzed the radio continuum map, which depicts a point-like source surrounded by a shallow nebula. From the flux of the point-like source, we derived a current mass-loss rate of $\dot M = (1.55\pm0.21)\times10^{-5}\, \mathrm{M}_{\sun}$ yr$^{-1}$. Finally, to better understand the complex circumstellar environment of AG Car, we put the newly detected ring in relation to the main nebula of dust and ionized gas. We discuss possible formation scenarios for the ring, namely, the accumulation of interstellar material due to the action of the stellar wind, the remnant of a close binary interaction or merger, and an equatorially enhanced mass-loss episode. If molecular gas formed in situ as a result of a mass eruption, it would account for at least a $30\%$ of the total mass ejected by AG Car. This detection adds a new piece to the puzzle of the complex mass-loss history of AG Car, providing new clues about the interplay between LBV stars and their surroundings.
\end{abstract}

% Select between one and six entries from the list of approved keywords.
% Don't make up new ones.

%%%%%%%%%%%%%%%%%%%%%%%%%%%%%%%%%%%%%%%%%%%%%%%%%%

%%%%%%%%%%%%%%%%% BODY OF PAPER %%%%%%%%%%%%%%%%%%

\section{Introduction}
\label{sec:introduction}

Over the course of their lifetime, massive stars have a great impact in the chemistry, structure and dynamics of the interstellar medium (ISM). This impact intensifies as they enter the luminous blue variable (LBV) phase, a brief ($\sim$ 10$^4$ years) and unstable post-main sequence stage, in which massive stars exhibit the highest mass-loss rates (up to 10$^{-3}$ $\mathrm{M}_{\sun}$ yr$^{-1}$). Such a copious mass loss occurs by virtue of dense and steady stellar winds and, sporadically, violent outbursts like the Great Eruption of $\eta$ Car in the 19th century. The interaction between this mass loss, interstellar material and different wind regimes shapes the closest circumstellar environment, leading to the formation of large multi-phase nebulae.

Nebulae around LBV stars (LBVNs) therefore emerge as valuable laboratories to understand the feedback mechanisms between the parent stars and their environment. The multi-wavelength study of this circumstellar material is a crucial tool to trace back the mass-loss record of these sources throughout their different evolutionary phases (e.g. \citealt{2005A&A...437L...1U, 2009ApJ...694..697U, 2010ApJ...718.1036U,2014A&A...562A..93C,2014MNRAS.440.1391A,2017MNRAS.465.4147B}). However, most of the research so far has focused on the dust and ionised gas content of LBVNs, paying little attention to the missing piece of the picture: the molecular gas.

In recent years, carbon monoxide and a handful of other simple molecular species have been detected in the surroundings of several candidate and confirmed LBVs, such as G79.29+0.46 (\citealt{2008ApJ...681..355R}, \citealt{2014A&A...564A..21R}), [GKF2010] MN101 (=MGE 042.0787+00.5084, \citealt{2019MNRAS.482.1651B}), and the well known $\eta$ Car (\citealt{2001ApJ...551L.101S}, \citealt{2012ApJ...749L...4L}, \citealt{2016ApJ...833...48L}, \citealt{2018MNRAS.474.4988S}, \citealt{2019MNRAS.490.1570B}, \citealt{2020ApJ...892L..23M}, \citealt{2020MNRAS.tmp.3075G}). All these successful detections demonstrate that, provided the adequate physical conditions, conspicuous amounts of molecular gas can arise and survive for some time in the hostile outskirts of these stars. By analyzing the emission at mm- and sub-mm wavelengths from rotational transitions of CO and other species, one can obtain valuable kinematic information that allows for precisely reconstructing previous mass-loss events and place constraints on the timescales of the observed structures. Therefore, the very existence of these molecules opens a complementary window to learn about the mass-loss phenomena in these hot, massive stars, beyond what is visible at other wavelengths. In addition, the determination of the relative chemical abundances and isotopic ratios in the molecular gas provides valuable snapshots of the stellar chemistry.

Among the scant LBV family, AG Car stands out as one of the most luminous members (L $\sim1.5\times10^6$  $\mathrm{L}_{\sun}$, \citealt{2011ApJ...736...46G}). Located in the far side of the Carina arm, the literature establishes a canonical distance of $6\pm1$ kpc \citep{1989A&A...218L..17H}. Revisions on this value were recently made by \cite{2017AJ....153..125S} and \cite{2019MNRAS.488.1760S} based on Gaia parallaxes. They concluded that AG Car may be 20\% closer, at a distance of $4.7^{+1.2}_{-0.8}$ kpc, which is consistent, within the uncertainties, with the previous estimate. Hereafter in this work, we adopt the canonical distance of 6 kpc. 

AG Car has been widely observed across the whole electromagnetic spectrum for decades. The star is surrounded by a slightly bipolar ring nebula of about $30\times40$ arcsec that was first reported by \cite{1950MNRAS.110..524T}. Further dynamic studies of the nebula in the light of H$\alpha$, [\ion{N}{ii}] and [\ion{S}{ii}] by \cite{1977MNRAS.180...95T} concluded that the nebula was a hollow shell with hints of bipolarity, expanding at $\sim$50 km s$^{-1}$. \cite{1989A&A...218L..17H} determined a kinematic age for the nebula of $\sim10^4$ years, consistent with the duration of the LBV phase. Broad-band optical continuum observations by \cite{1989ApJ...341L..83P} revealed a helical jet-like structure in the NE-SW direction, apparently arising from the star, which gave rise to questions about the possible binarity of AG Car. However, no signature of a companion star was detected \citep{1992ApJ...398..621N}. \cite{1991IAUS..143..385S} and \cite{1992ApJ...398..621N} revisited the dynamics of the nebula, establishing an expansion velocity of 70 km s$^{-1}$ and identifying a bipolar outflow at 83 km s$^{-1}$ distorting the NE side of the shell. Later, ATCA observations at 3 and 6 cm by \cite{2002MNRAS.330...63D} detected radio continuum emission arising from the star and the shell, with a spectral index consistent with thermal radiation from ionised gas. Finally, the dust content of the nebula was thoroughly investigated by \cite{2000A&A...356..501V} and \cite{2015A&A...578A.108V} (hereafter VN15) by means of infrared imaging and spectroscopy. The latter estimated a total nebular mass of $\sim$15 $\mathrm{M}_{\sun}$ and proposed a mass eruption with a kinematic age of $1.7\times10^4$ yr as the origin of the structure.

None the less, what makes AG Car particularly interesting is its extreme variability. AG Car exhibits yearly micro-variations of 0.1--0.5 mag superimposed to a more than a decade-long S Dor cycle, with visual changes of $\sim$ 2 mag between the hot and cool states  \footnote{https://www.aavso.org} \citep{1994PASP..106.1025H, 2001A&A...375...54S, 2003AAS...202.1303S}. For that reason, AG Car is an excellent laboratory to learn about the formation and destruction of molecules in a changing environment, in terms of physical conditions and time scales.

The first attempts to detect molecular gas associated with AG Car were made by \cite{2002AJ....124.2920N} (hereafter N02), targeting the CO $J=1\rightarrow0$ and $J=2\rightarrow1$ lines with the SEST telescope. With a series of single-dish pointings, they coarsely sampled the region around the star, with beam sizes of 45 and 23 arcsec and spacings of 45 and 12 arcsec respectively. Their spectra displayed multiple narrow velocity components, likely related to intervening sheets of molecular gas, but also a broad component centred at 26 km s$^{-1}$. The authors interpreted this broad component, which presented a pseudo-gaussian profile, as arising from a circumstellar expanding envelope or disk, for which they estimated a minimum mass of 2.8 $\mathrm{M}_{\sun}$. Despite this promising result, the region was never properly imaged at higher resolution to confirm the existence of such a structure.

In this work, we report APEX and ALMA observations of CO and $^{13}$CO towards AG Car, confirming the detection of a molecular ring surrounding the star. The paper is structured as follows: in Sect. 2 we summarize the molecular line observations; in Sect. 3 we describe the main findings; in Sect. 4 we analyze the morpho-kinematic features of the structure, derive the physical parameters of the gas and present a kinematic model. In the same section, we also discuss possible formation mechanisms. In Sect. 5 we put together the available data into a single, unified picture of the mass-loss record of AG Car; and finally, in Sect. 6 we present the conclusions of this study and lay the groundwork for further research on AG Car.

\section{Observations}

\subsection{APEX observations}

We observed AG Car with the Atacama Pathfinder EXperiment (APEX) telescope, located at Llano de Chajnantor (Chile). Observations took place on 2014, September 23, as part of the program E-094.D-0598A (P.I: G. Umana). The front-end used was the APEX2 receiver of the Swedish Heterodyne Facility Instrument (SHEFI, \citealt{2008A&A...490.1157V}), targeting the CO $J=3\rightarrow2$ transition at 345.796 GHz. The eXtended Fast Fourier Transform Spectrometer (XFFTS, \citealt{2012A&A...542L...3K}) provided an instantaneous bandwidth of 2.5 GHz with 32768 channels, resulting in an effective velocity resolution of about 0.07 km s$^{-1}$ at the observing frequency.

An On-The-Fly map of $100\times100$ arcsec around the source was done in Total Power mode, using the off position $\alpha =10^\mathrm{h}49^\mathrm{m}23^\mathrm{s}.3, \,\delta=-60^\circ 58\arcmin 47\arcsec.7$ (J2000). At the rest frequency of the line, the APEX primary beam is 19.2 arcsec. Observations were performed under average weather conditions (1--2 mm of precipitable water vapour). Calibration was done using Mars and X TrA, and pointing and focus were checked regularly during the night.

%\subsubsection{Data reduction}

The raw spectra produced by the APEX standard pipeline was reduced using the \textsc{gildas}\footnote{http://www.iram.fr/IRAMFR/GILDAS} software package. This process involved the removal of bad scans, a linear baseline fitting and a velocity smoothing to a final resolution of 1 km s$^{-1}$ to improve rms. Conversion from the original antenna temperatures ($T_A^*$) to main-beam brightness temperature (T$_\mathrm{mb}$) was done by correcting for the forward-hemisphere efficiency and the beam efficiency of the antenna, so that

\begin{equation}
    T_\mathrm{mb} = T_\mathrm{A}^*\frac{F_\mathrm{eff}}{B_\mathrm{eff}}.
\end{equation}

For the APEX telescope, $F_\mathrm{eff}$ = 0.97 and $B_\mathrm{eff}$ = 0.73 at 345 GHz. The typical rms noise per 1 km s$^{-1}$ channel in the final cube is 50 mK.

\begin{table}
\begin{tabular}{@{}lllll@{}}
\toprule
Line    & $\nu_0$ [GHz] & Telescope & $\theta$ [''] \\ \midrule
\multirow{2}{*}{CO $J=2\rightarrow1$}   & \multirow{2}{*}{230.538}  & ACA    & 7.0$\times$5.2  \\
                           &                                        & TP     & 28.2 \\
\multirow{2}{*}{$^{13}$CO $J=2\rightarrow1$} & \multirow{2}{*}{220.398} & ACA        & 7.2$\times$6.0   \\
                           &                      & TP         & 29.5   \\
CO $J=3\rightarrow2$                    & 345.795                  & APEX       & 19.2  \\ \bottomrule
\end{tabular}
\caption{Observational parameters.}
\label{tab:parameters}
\end{table}

\subsection{ALMA observations}

Follow up observations of AG Car were conducted on 2019, October 2 and 6, with the ALMA 7-m and total power arrays, respectively, in the context of the program 2019.1.01056.S (P.I: L. Cerrigone). It was mapped the spatial distribution of the rotational $J=2\rightarrow1$ transitions of CO and $^{13}$CO at 1.3 mm. The source was observed with three total-power antennas under excellent weather conditions (pwv $\sim$0.25 mm) and with ten 7-m antennas under good conditions (pwv $\sim$1.25 mm).

The spectral setup consisted of two spectral windows centred at 220.398 and 230.538 GHz plus two additional windows for continuum. The correlator was tuned to provide an instantaneous bandwidth of 1850 MHz in 2048 channels, with a moderate velocity resolution of about 1.3 km s$^{-1}$ for each line.

ACA data were calibrated with CASA v.5.6.1-8 and the ALMA pipeline version 42686 that comes with it. Calibrators J1047-6217 and J1058+0133 were used for bandpass and flux calibration. Cubes of line emission, with a characteristic beam of $7\times5$ arcsec, were constructed from the calibrated visibilities using the CLEAN algorithm after continuum-subtraction. CASA v5.6.1-8 and pipeline v.42686 were also used for the reduction of the single-dish data. The observations were performed in position-switching mode, with an off position chosen by the ALMA Observatory among known clean areas for Galactic sources. The off position was about 4 deg away from our target. Such a large angular distance between the target and the off position caused ripples in the spectra, which we did not attempt to remove in post-processing, since their amplitude is negligible with respect to the brightness of the spectral lines from our target. The flux calibration was based on Kelvin-to-Jansky factors estimated from calibration campaigns performed by the Observatory. At our frequencies the factors had a value of about 41.5 K/Jy.

Finally, to recover all the spatial scales and avoid missing flux, we combined the ACA and the Total Power (hereafter ALMA TP) cubes with the CASA \texttt{feather} task, which involved regridding the single-dish data and correcting it for the primary beam response. The feathered cubes recover 99.7\% of the single-dish flux.

The resulting products were then re-scaled from flux densities to brightness temperature using the standard equation

\begin{equation}
    T_\mathrm{mb} = 1.222\times10^6 \frac{S_\nu}{\nu^2 \theta_\mathrm{maj} \theta_\mathrm{min}},
\end{equation}

\noindent where $S_\nu$ denotes the flux density in Jy beam$^{-1}$, $\nu$ is the reference frequency in GHz, and $\theta_\mathrm{maj}$ and $\theta_\mathrm{min}$ are the major and minor axes of the beam in arcsec, respectively. The typical rms noise level per channel of 1.3 km s$^{-1}$ ranges from 5 to 10 mK. Similarly, the rms noise level of the continuum map is $\sim1.2$ mJy beam$^{-1}$.\\
\newline
A summary of the observing parameters is presented in Table \ref{tab:parameters}. Throughout this paper, we use the following conventions: (1) intensities of emission lines are expressed in $T_\mathrm{mb}$ scale; (2) intensities of continuum are expressed in flux density scale; (3) velocities refer to the local standard of rest frame (LSR); and (4) positions are offsets from the coordinates of AG Car, $\alpha =10^\mathrm{h}56^\mathrm{m}11^\mathrm{s}.5779, \,\delta=-60^\circ 27\arcmin 12\arcsec.8095$ (J2000) \citep{2018yCat.1345....0G}.

\section{Results}\label{sec:results}

\subsection{APEX sigle-dish data}\label{sec:results:apex}
\begin{figure}
 \includegraphics[width=\columnwidth]{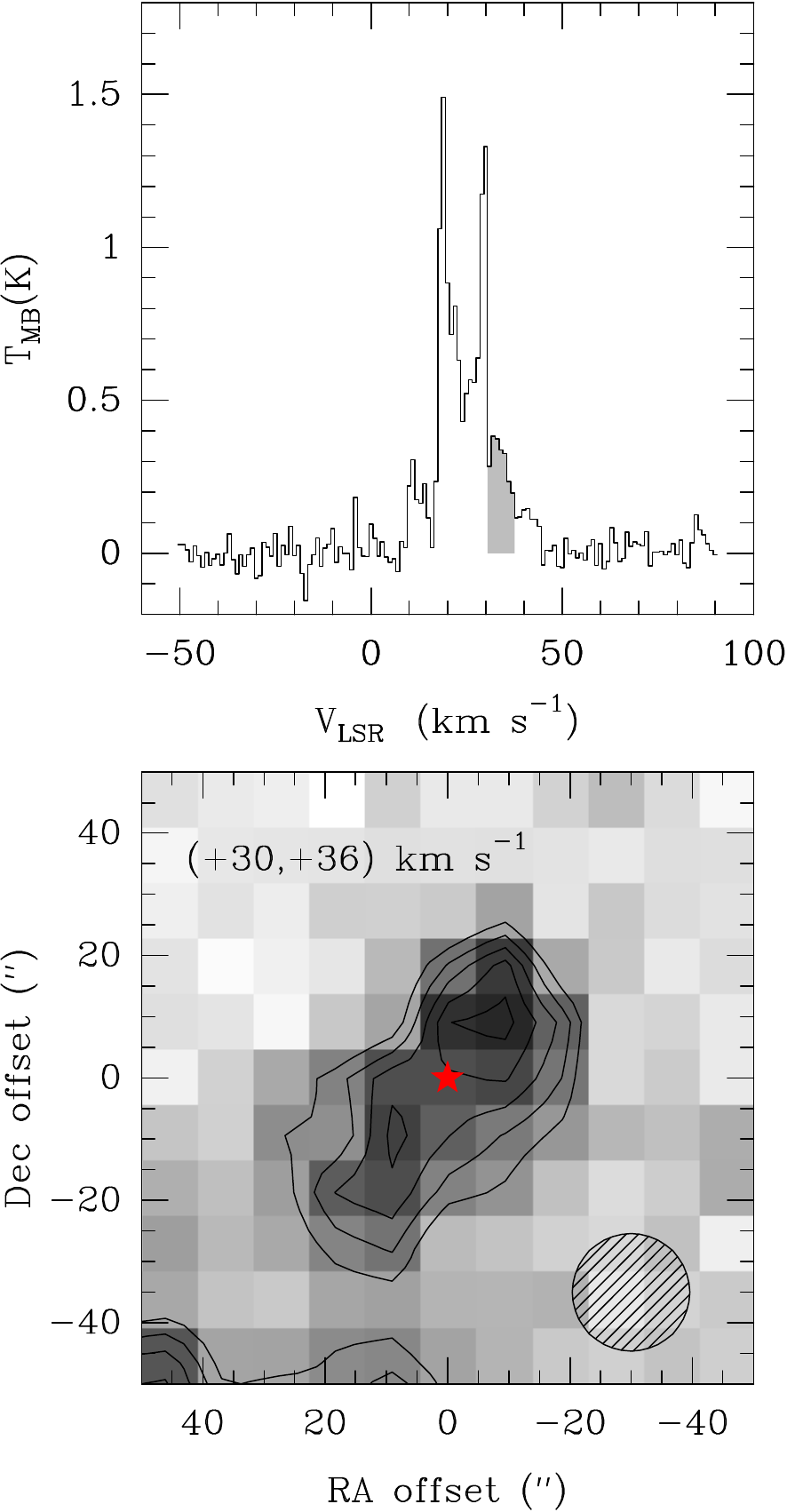}
 \caption{APEX CO $J=3\rightarrow2$ emission towards AG Car. Top panel: spectrum averaged in the 100$\times$100 arcsec field. The velocity range integrated in the bottom panel is filled in grey. Bottom panel: velocity-integrated intensity map from 30 to 36 km s$^{-1}$, corresponding to the filled area in the top panel. Contours start at 2.5 K km s$^{-1}$ in steps of 0.5 K km s$^{-1}$. The red marker indicates the position of the star, and the hatched circle in the bottom right corner represents the APEX primary beam size.}
 \label{fig:apex}
\end{figure}

Fig. \ref{fig:apex} shows the averaged APEX CO $J=3\rightarrow2$ spectra in the field of AG Car. Most of the emission arises in the velocity range (0, 50) km s$^{-1}$. We identify multiple velocity components within this range, coincident with the ones reported by N02 in the CO $J=1\rightarrow0$ and CO $J=2\rightarrow1$ transitions: two narrow, dominant components at 18.5 and 29.7 km s$^{-1}$ plus additional minor components at 11, 18.5, 21.5 and 29.5 km s$^{-1}$. N02 attributed these components to background or foreground contamination from ambient gas moving perpendicularly to the line of sight. We also detect a broad component at 26.5 km s$^{-1}$, the same one that N02 tentatively associated with the star.

Careful inspection of the spatial distribution of each component shows that all the narrow lines are clumpy and somewhat extended, mainly arising from the southeast and the west. The same applies to the broad component, which is particularly intense towards the centre and the south of the map. However, we note that the 29.5 km s$^{-1}$ narrow line presents a 'shoulder-like' feature consistent with a partially overlapping line at a slightly higher velocity. This component was indeed marginally detected in N02's spectra, without clear hints of any particular spatial distribution, probably being just a high-velocity wing of the broad component at 26.5 km s$^{-1}$.

This newly identified component, centred at 32.5 km s$^{-1}$, has a velocity extension of about 6 km s$^{-1}$ and is significantly weaker than the others. We attempted to isolate the line by removing the neighbouring components through a gaussian fitting, but this resulted in the introduction of multiple artefacts that altered the line shape. None the less, integration of the approximate velocity range of the line, i.e. at least 30 to 36 km s$^{-1}$, represented in grey in the figure, reveals the existence of an elongated structure towards the centre of the field (Fig. \ref{fig:apex}, bottom panel). Despite being possibly contaminated by the adjacent narrow component, the structure exhibits a certain degree of bipolar symmetry with respect to AG Car, with two 'lobes' along the SE-NW axis. None of the other ambient components shows a comparable spatial distribution.

\subsection{ACA interferometric data}\label{sec:results-aca}

\subsubsection{CO and $^{13}$CO emission}
\label{sec:results:aca}

\begin{figure}
 \includegraphics[width=\columnwidth]{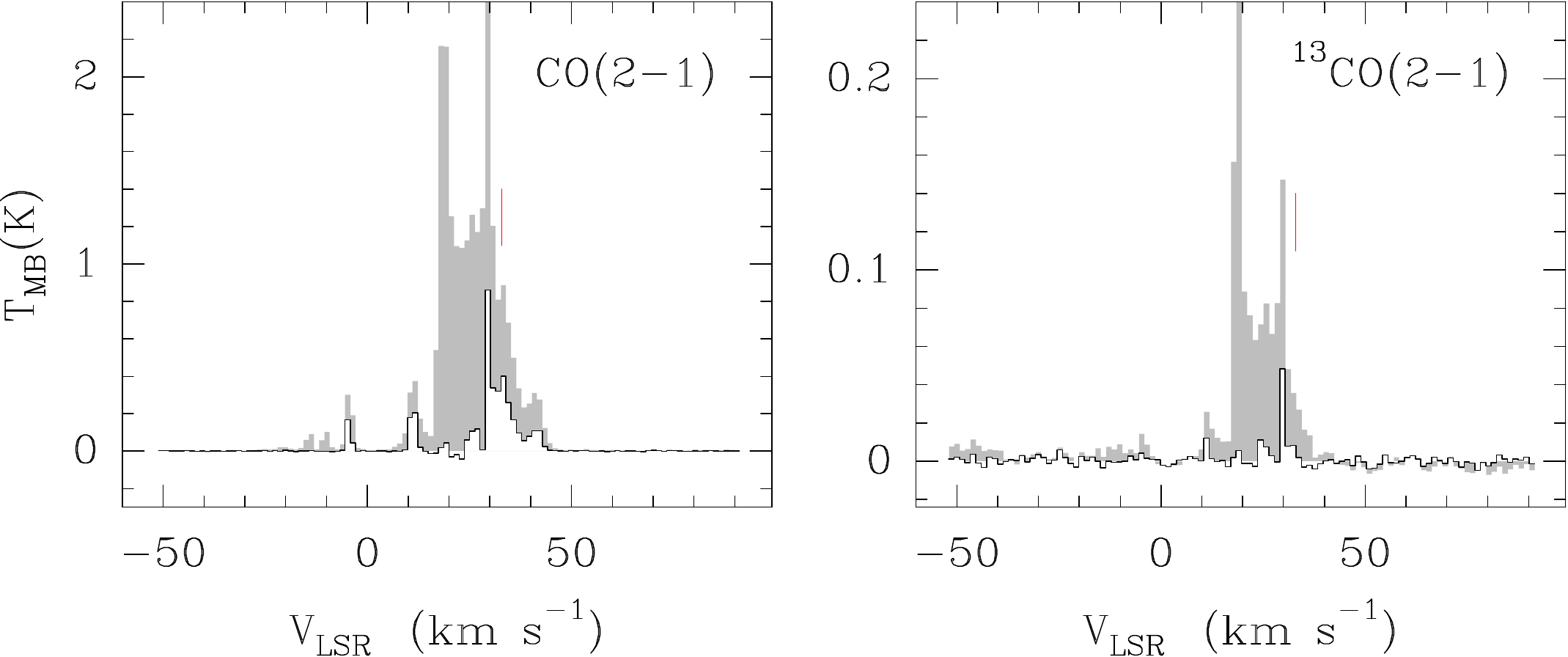}
 \caption{Interferometric filtering towards AG Car. Left panel: CO $J=2\rightarrow1$. Right panel: $^{13}$CO $J=2\rightarrow1$. In each panel, feathered ACA+TP (filled) and ACA only spectra (black), averaged within a radius of 20 arcsec from the star, are displayed in the same temperature scale. The red lines indicate the central velocity of the component likely associated with the star.}
 \label{fig:spec-aca}
\end{figure}

The feathered CO $J=2\rightarrow1$ and $^{13}$ CO $J=2\rightarrow1$ spectra show essentially the same features visible in the CO $J=3\rightarrow2$ APEX data. Fig. \ref{fig:spec-aca} compares the feathered (filled) and the ACA-only (black) spectra for the two observed lines. The interferometer filtered out most of the emission at $v_\mathrm{LSR} <$ 30 km s$^{-1}$, including the broad line at 26.5 km s$^{-1}$. Considering the ACA maximum recoverable scale is 28 arcsec at 230 GHz (our shortest baseline was $\sim8.9$ m), these components necessarily correspond to extended, large scale emission, as previously discussed. On the other hand, a significant fraction of the flux in the velocity range $(+30, +40)$ km s$^{-1}$ is preserved, suggesting that this emission is considerably more compact than the ambient material. From the peak temperature, we estimate that ACA recovers $\sim$50\% of the single-dish flux in this range.

\begin{figure}
 \includegraphics[width=\columnwidth]{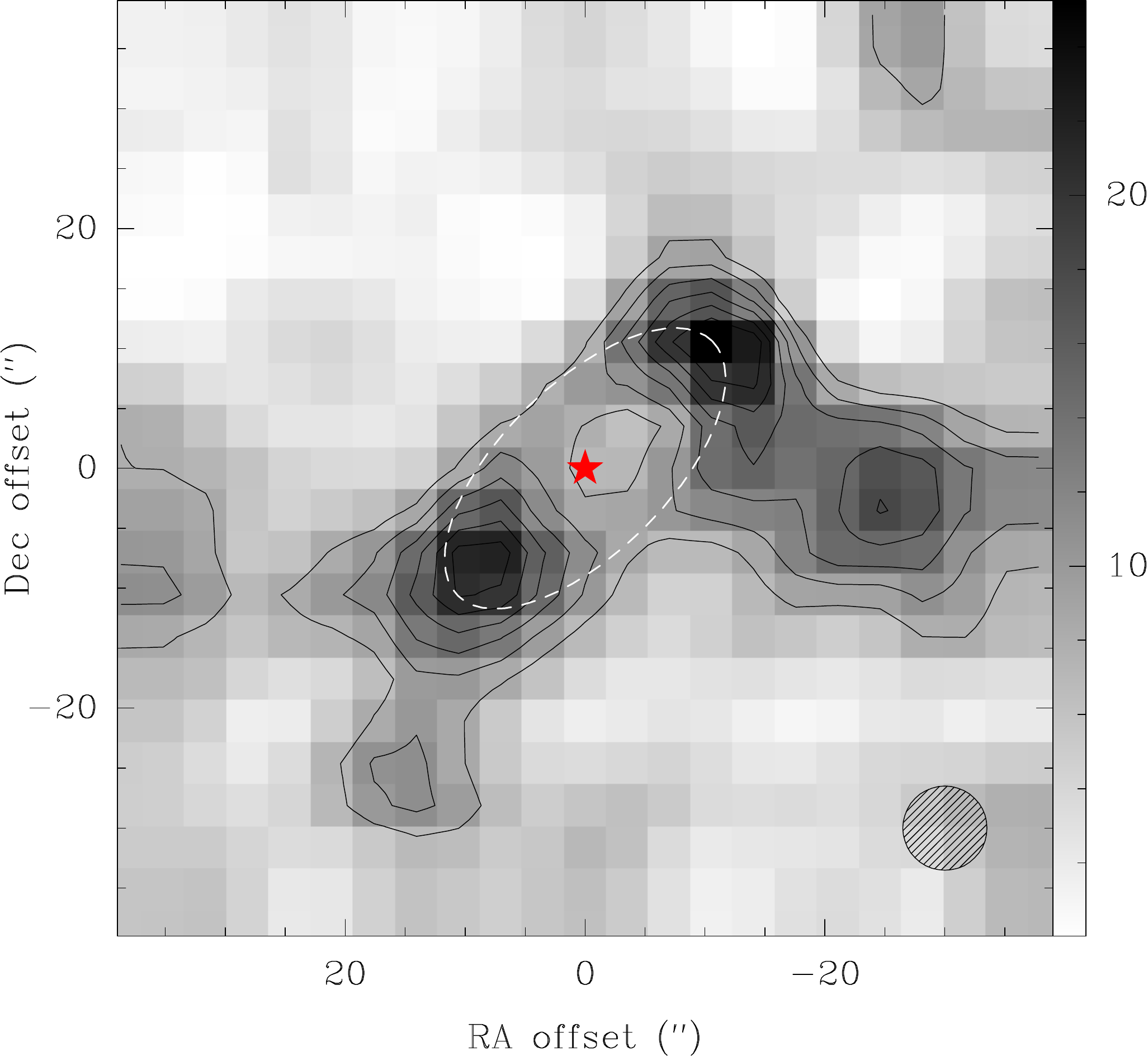}
 \caption{ACA+TP velocity-integrated intensity map of CO $J=2\rightarrow1$ towards AG Car, in the velocity range ($+29$,$+36$) km s$^{-1}$. Contours start at 8 K km s$^{-1}$ in steps of 2.5 K km s$^{-1}$. The red marker indicates the position of the star, and the hatched circle in the bottom right corner represents the ACA synthesized beam size. The dashed ellipse represents the ellipse used for the fitting.}
 \label{fig:aca}
\end{figure}

Fig. \ref{fig:aca} shows the integrated intensity map of the CO $J=2\rightarrow1$ line in the range ($+$29, $+$36 km s$^{-1}$). ACA's synthesized beam (of $\sim$7 arcsec) improves angular resolution with respect to APEX by a factor of $\sim$3. Such improvement allows us to resolve the emission, revealing an elliptical structure enclosing AG Car and apparently detached from the star. Two remarkably symmetric 'lobes' dominate the emission, southeast and northwest of the star, peaking at (10\arcsec,--10\arcsec) and (--10\arcsec, 10\arcsec) respectively. The lobes appear connected by two fainter arcs. A large clump at position (--25\arcsec, --5\arcsec) disrupts the structure in its westernmost side. This clump is related to residuals from the narrow component at 29.5 km s$^{-1}$ that are visible in the ACA spectra (perhaps the densest part of an extended cloud). The S/N ratio across the ring ranges from 6 $\sigma$ to 14 $\sigma$, which means that even the faintest parts are genuine. By fitting an ellipse to the lobes, we derive an approximate angular size of 35$\times$15 arcsec, with a position angle of 135 degrees east of north. The orientation and size of the ellipse match perfectly the NW-SE elongation of the infrared shell reported by \cite{2000A&A...356..501V} from continuum and [\ion{Ne}{ii}] imagery. In the $^{13}$CO $J=2\rightarrow1$ integrated intensity map (not shown), only the lobes of the structure are visible, with an S/N ratio slightly above 5 $\sigma$. The non-detection of fainter arcs is likely due to the limited sensitivity.

\subsubsection{Radio continuum emission}\label{sec:results:continuum}

\begin{figure}
 \includegraphics[width=\columnwidth]{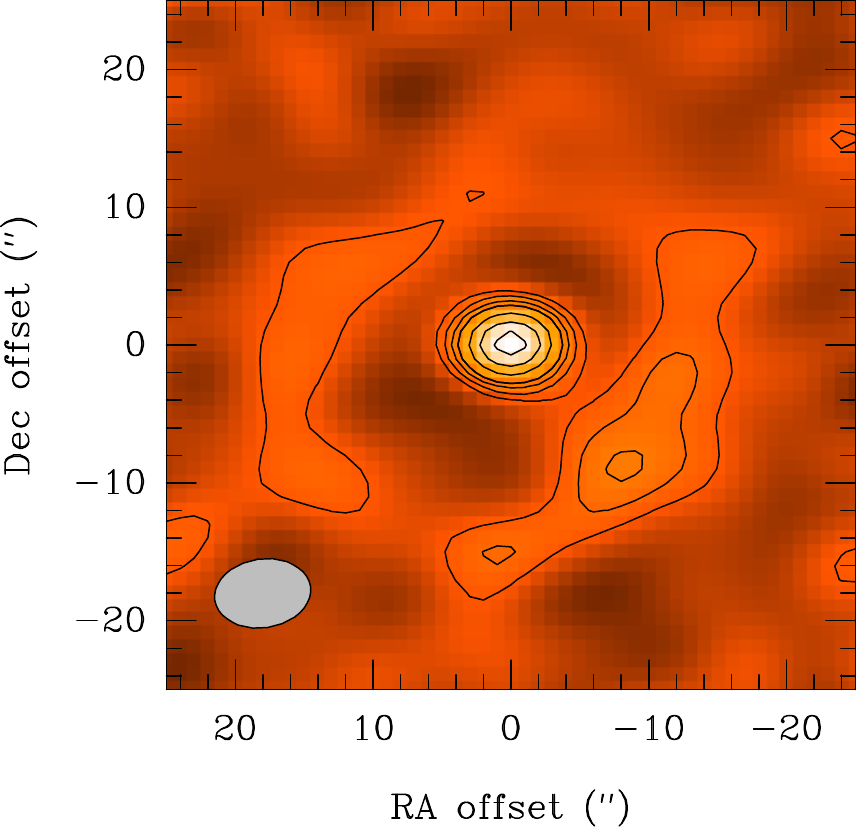}
 \caption{ACA 225 GHz radio continuum map of AG Car. Contours start at 2.5 mJy, with steps of 2.5 mJy. The synthesized beam is shown in the bottom left corner.}
 \label{fig:aca-cont}
\end{figure}

Radio continuum emission traces the ionised gas in the nebula of AG Car. Fig. \ref{fig:aca-cont} shows the ACA radio continuum map of AG Car at 225 GHz. Emission is dominated by an unresolved compact source in the center of the field, presumably related to the star. Weak emission arising from the brightest parts of the optical nebula are visible as well above the $3\sigma$ level.

The radio morphology is slightly different from the 5.5 and 8.8 GHz ATCA maps presented by \cite{2002MNRAS.330...63D}, which depict a compact source surrounded by a bright and very clumpy detached nebula. Note, though, that ATCA data has a higher spatial resolution (1 arcsec at 8.5 GHz) and provides a more complete \textit{uv} coverage.

We fitted a two-dimensional gaussian to the compact source, measuring a flux density of $30.2\pm2.2$ mJy. The uncertainty is the quadratic sum of the rms noise in the map and a calibration error of 5\%. The resulting value does not match the expected flux density obtained by extrapolating the spectral index $\alpha=-0.1$ reported by \citealt{2002MNRAS.330...63D}, which is 0.89 mJy. Such discrepancy implies that either the spectral index is variable, and thus the current physical conditions in the star are quite different from those in the 1994--1996 period (when the ATCA data were gathered), or that another emitting component is superposed at mm-wavelengths, such as thermal dust close to the star, effectively pushing the spectral index toward more positive values. Finally, if we ignore the 25 year gap between the observations and combine the fluxes, we obtain a spectral index $\alpha\sim0.8$, compatible with a typical stellar wind.

We determined a nebular flux of $40\pm15$ mJy after subtracting the contribution of the point source. Again, this value does not match the extrapolation of the $-0.1$ spectral index reported by \cite{2002MNRAS.330...63D}, which yields a slightly higher flux. We note, though, that the LAS (largest angular scale) of ACA at 225 GHz is about 1/3 of the LAS of ATCA 750B configuration at 5.5 GHz, so it is very likely that we are missing a fraction of the extended flux.

\section{Linking the molecular ring to AG Car}\label{sec:discussion}

\subsection{Morphology and kinematics}\label{sec:morph-kin}

\begin{figure*}
 \includegraphics[width=\textwidth]{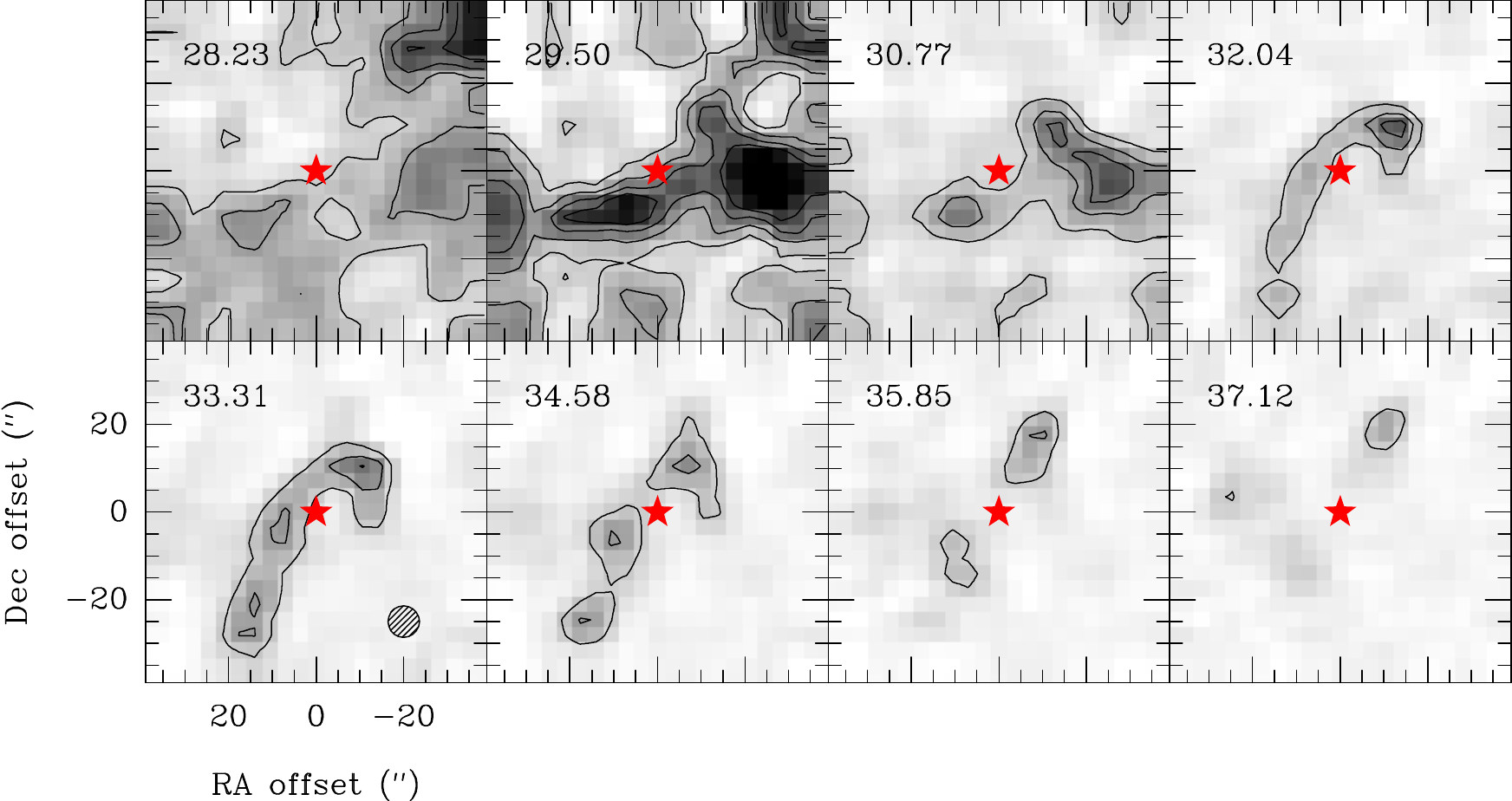}
 \caption{Chanel maps of CO $J=2\rightarrow1$ (ACA+TP). The velocity in km s$^{-1}$ is indicated in the top left corner of each panel. Contours start at $\sim3\sigma$ in steps of 3$\sigma$. The red marker indicates the position of the star, and the hatched circle in the in the bottom left panel represents the approximate ACA synthesized beam.}
 \label{fig:chan}
\end{figure*}

To establish a physical connection between the CO structure and AG Car, we need to analyze the morphology and kinematics of the gas and put them in context with the surrounding ISM. AG Car lies in projection amid the massive star clusters Tr 14 and Tr 16 and the large Car OB2 association. Toward these structures, i.e. from $l=287\degr$ to $290\degr$, several giant molecular clouds are observed. Most of them arise at negative systemic velocities, comprised between $-10$ and $-30$ km s$^{-1}$, corresponding to distances of 2--3 kpc on the near side of the Carina arm \citep{1985ApJ...290L..15C}. These clouds are dense, clumpy and continuously eroded by intense UV radiation from hundreds of newborn OB stars (e.g. \citealt{1998A&A...332.1025R, 2002MNRAS.331...85R,2018A&A...618A..53W}). On the contrary, the emission in the field of AG Car appears at $v_\mathrm{LSR} >$10 km s$^{-1}$. In this direction, these positive velocities are expected for gas on the far side of Carina \citep{1987ApJ...315..122G}, which agrees with the distance of 6 kpc assumed for the star.

\begin{figure}
 \includegraphics[width=\columnwidth]{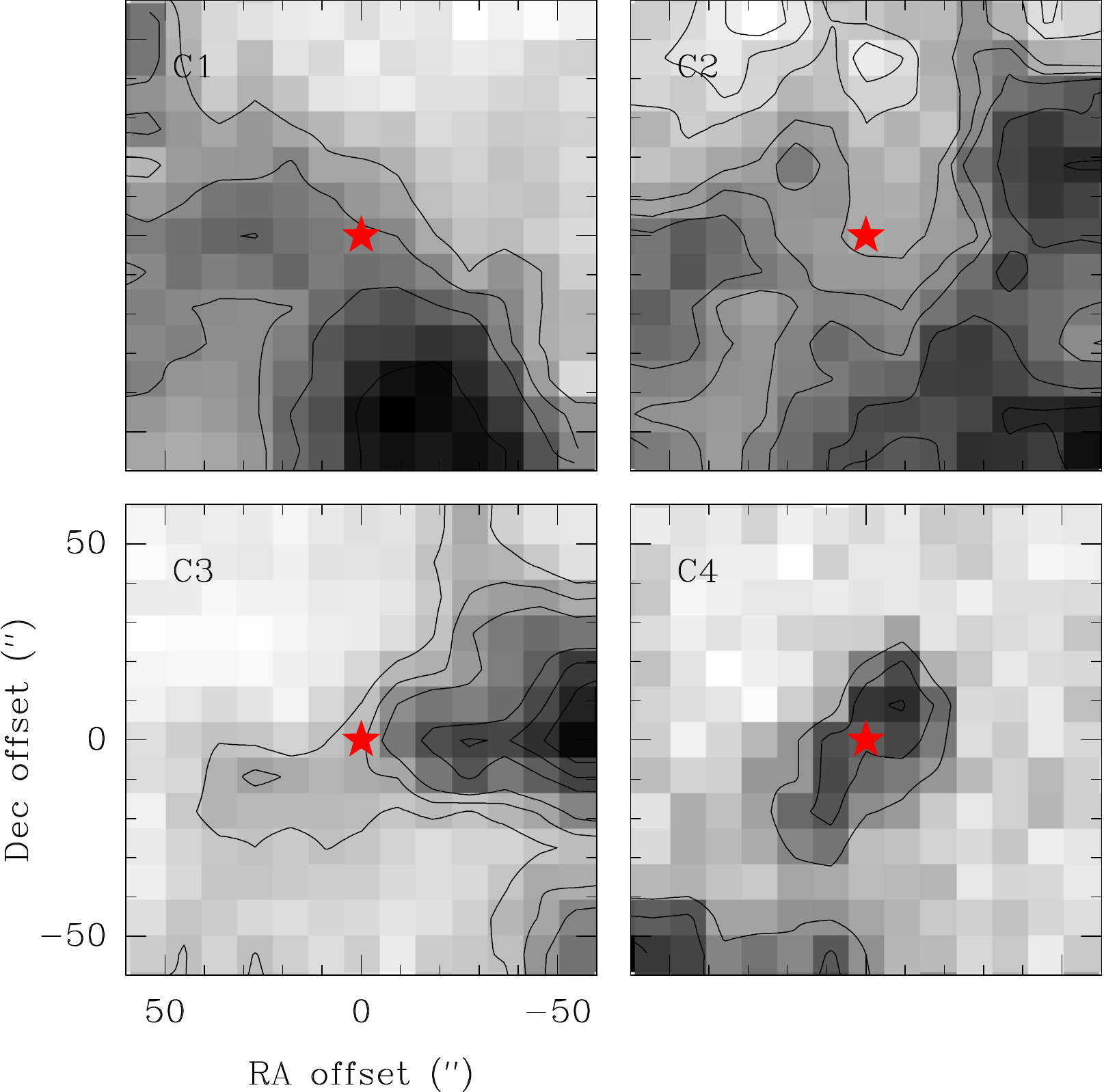}
 \caption{Spatial distribution of the main components in the field of AG Car. APEX CO $J=3\rightarrow2$ velocity-integrated intensity maps of components C1 (from 17.5 to 19.7 km s$^{-1}$), C2 (from 19.3 to 28.2 km s$^{-1}$), C3 (from 28.4 to 30.8 km s$^{-1}$) and C4 (from 30 to 36 km s$^{-1}$, same as Fig. \ref{fig:apex}). The colour scale is relative to the peak intensity of each map. Contours at 2.5, 3.5, 4.5, 5.5, 6.5, 8.5, 10.5 and 12.5 K km s$^{-1}$. The red marker indicates the position of the star.}
 \label{fig:contaminants}
\end{figure}

A closer look at the distribution of the emission in the field of AG Car highlights the differences between the central CO structure and the other velocity components. Fig. \ref{fig:contaminants} shows the APEX CO $J=3\rightarrow2$ velocity-integrated intensity maps of the dominant lines, i.e. the two narrow lines at 18.6 and 29.6 km s$^{-1}$ and the broad component at $26.5$ km s$^{-1}$ that N02 tentatively linked with the star (labelled C1, C2 and C3 in order of increasing velocity). As mentioned earlier in Sect. \ref{sec:results:apex}, none of the three components seem to be morphologically associated with the star, unlike the component around 32.5 km s$^{-1}$ (hereafter C4), shown again for reference. The "arm" seen in component C3, extending over the source to the East, is due to contamination by the low velocity wing of component C4 (the two lines overlap significantly).

Most of the emission from C1, C2 and C3 is resolved out by the interferometer, preserving the small-scale features of component C4. As seen in the ACA map, the spatial distribution of C4 is consistent with a circumstellar ring or torus seen at a certain inclination angle. Assuming that such structure is perfectly axisymmetric, we can make an elliptical fit and use the axial ratio $R_\mathrm{ab}$ to estimate the viewing angle, such that $\theta_i = \arctan(R_\mathrm{ab})$. This method yields an angle of $\theta_i = 68\pm3$ degrees, which implies that the ring is viewed nearly edge-on, with its semimajor axis following the direction in which the infrared shell is slightly elongated \citep{2000A&A...356..501V}. The deprojected characteristic radius of the structure is thus 15 arcsec, which is equivalent to $\sim$0.4 pc at a distance of 6 kpc. This physical size is in good agreement with the inner radius of the infrared dusty shell derived by VN15.

The ring extends from 29 to 36 km s$^{-1}$, with a central velocity of 32.5 km s$^{-1}$. In the literature, it is possible to find estimates for the radial velocity of AG Car spanning from 0 to 10 km s$^{-1}$ (e.g. \citealt{1982A&A...112..111W, 1989A&A...218L..17H, 2001A&A...375...54S, 2009ApJ...698.1698G}). Such discrepancy is the consequence of the inherent uncertainty in determining systemic velocities for LBV stars, whose variable stellar winds and circumstellar envelopes often contaminate spectroscopic measurements. Contrarily, the velocity of the CO structure does not depend on any stellar properties. Therefore, in our analysis, we consider a systemic velocity of 32.5 km s$^{-1}$. This velocity is compatible with the distance assuming small departures from galactic rotation.

Fig. \ref{fig:chan} shows the channel maps of the CO $J=2\rightarrow1$ emission in the velocity range of interest. We note that the moderate velocity resolution of our data, of just 1.3 km s$^{-1}$, does not allow us to perform a detailed analysis of the observed kinematic structure, which is sampled by only seven channels. The first two channels, corresponding to $v_\mathrm{LSR}$ = 28.2 and 29.5 km s$^{-1}$, appear highly contaminated, with a prominent clump towards the west particularly bright in the latter. As discussed in Sect. \ref{sec:results-aca}, this clump is likely associated with the narrow component at 29.5 km s$^{-1}$: the map of C3 presents relative maximum near position (--25\arcsec, --5\arcsec), as seen in Fig. \ref{fig:contaminants}. The rest of the channels are mostly devoid from contamination, except for small faint clumps towards the south. Overall, the emission moves clockwise around the star, from the southwest to the northwest, as velocity increases. The lobes are present across all the channels, with the southeast one being slightly more prominent in the blueshifted channels and vice-versa. Contrarily, the arcs that connect the lobes are only visible in the central channels: opening towards the northeast at 29.5 and 30.8 km s$^{-1}$, and towards the southwest at 33.3 and 34.6 km s$^{-1}$. 

The velocity gradients observed along the minor and major axes of the structure are mainly compatible with gas moving radially. Two physical scenarios are possible: (1) an accretion disc of gas falling onto the star; or (2), an expanding ring, composed of stellar ejecta or compressed interstellar gas. From an evolutionary perspective, the accretion disk scenario is highly improbable. Infalling disks around massive stars are expected only in pre-MS stages, and the evolved status of AG Car is more than confirmed. On the other hand, an expanding ring is a much more plausible hypothesis. In this case, the orientation is constrained by the observed velocity gradient: the southeastern part, blueshifted, is the approaching near side, and the northwestern part, redshifted, is then the receding far side. Correcting for the inclination, we estimate an expansion velocity of $v_\mathrm{exp}=3.5\pm0.5$ km s$^{-1}$. This value is surprisingly low, considering that typical velocities of LBV winds range from 50 to 100 km s$^{-1}$. For a characteristic radius of 0.4 pc, the kinematic age of the ring would be $\sim10^5$ years. This age must be seen as a strict upper limit, since the structure may be slowing down due to the interaction with the surrounding medium. Yet, the resemblance to the CO torus found in MN101 is noteworthy, having comparable sizes, expansion velocities and dynamic time-scales \citep{2019MNRAS.482.1651B}. These slow structures could well be the ageing relatives of the 200-yr-old, disrupted molecular torus in $\eta$ Car \citep{2018MNRAS.474.4988S}, which is more compact ($\sim$4000 au) and chemically rich  \citep{2019MNRAS.482.1651B, 2020MNRAS.tmp.3075G}. The idea of a common formation mechanism for these rings is worth to be studied in more detail. We explore the possible origins of the CO ring in Sect. \ref{sec:origin}.

\subsection{Physical parameters of the gas}

If the gas arises from the star --i.e. formed from wind or ejecta--, its physical parameters should reflect substantial differences with respect to the ISM. Line ratios are useful diagnostic tools to roughly estimate these parameters. We have three available lines to work with: two transitions of CO plus an additional line of $^{13}$CO. However, the analysis is restricted to the single-dish data, i.e. APEX and ALMA TP, since we lack interferometric CO $J=3\rightarrow2$ observations. Thus, we can only provide average values for the structure.

\begin{figure}
 \includegraphics[width=\columnwidth]{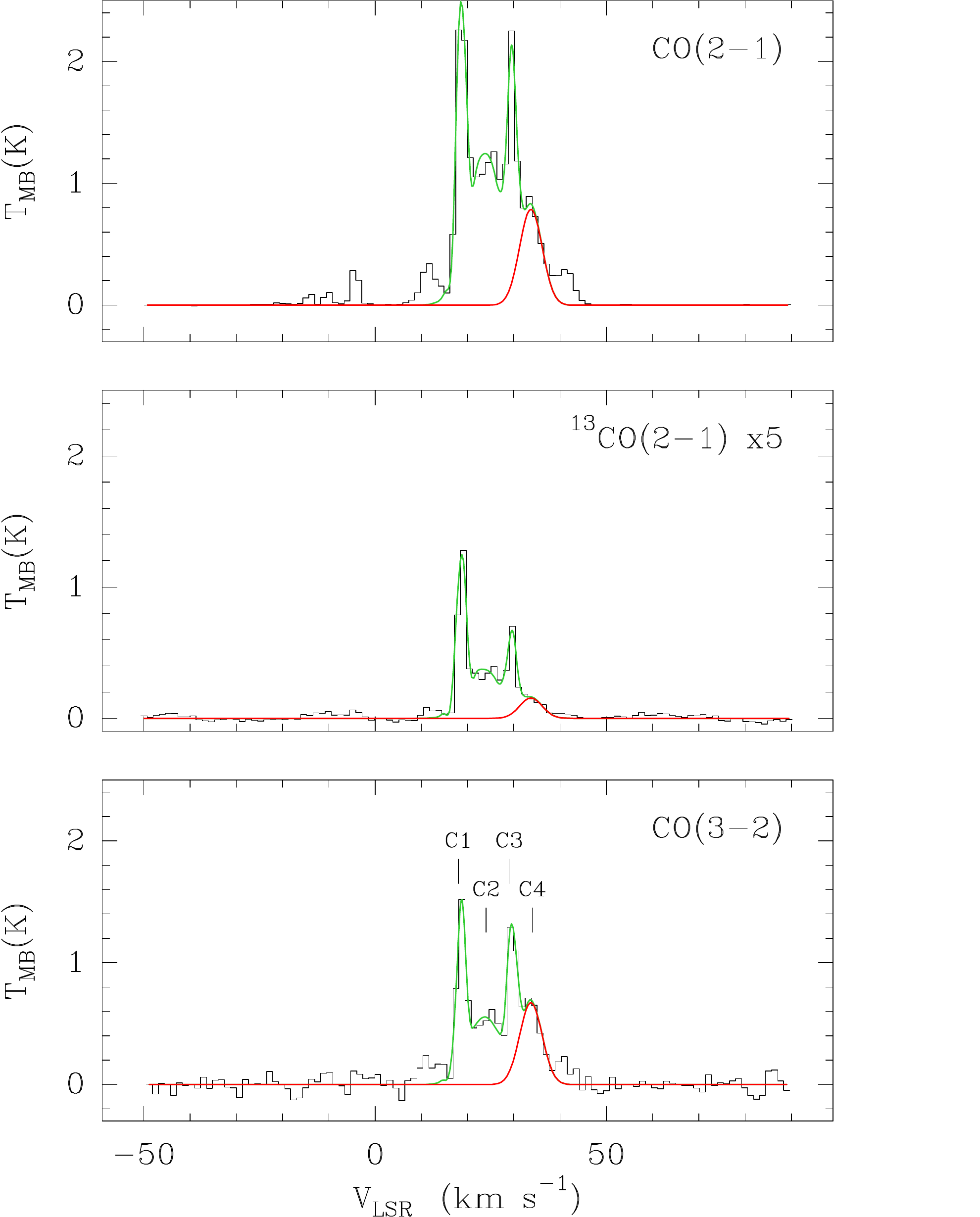}
 \caption{Fitting of the four dominant velocity components in the single-dish spectra. From top to bottom, CO $J=2\rightarrow1$, $^{13}$CO $J=2\rightarrow1$ (scaled by a factor of 5) and CO $J=3\rightarrow2$. The green line indicates the sum of the gaussian fittings of the four components, and the red line highlights the fitting for component C4. The components are labelled in the bottom panel for reference.}
 \label{fig:spec}
\end{figure}

We first computed beam-averaged spectra for the three transitions, centred in the stellar position. Then, we smoothed the resulting averaged spectra to a common velocity resolution of 1.3 km s$^{-1}$, and we attempted a gaussian fitting on components C1...C4 with \textsc{class}, as shown in Fig. \ref{fig:spec}. The results of the fitting are compiled in Tab. \ref{tab:line-params}. In some cases, the fitting was rather problematic due to significant line blending, derived from the limited velocity resolution. Therefore, we imposed the condition that every velocity component C$_i$ should have a same FWHM in each of the three lines, so that the derived line ratios were consistent. %According to our interpretation components C1 to C3 correspond to ambient material, whilst component C4 corresponds to the ring.

To calculate the line ratios, we used the line intensities integrated over the FWHM of each component, formally expressed as $W = \int{T_\mathrm{mb}\mathrm{d}v}$. The intrinsic uncertainty associated with this magnitude, $\sigma_W$, is given by

\begin{equation}
    \sigma_W = \mathrm{rms} \times dv \times \sqrt{N}
\end{equation}

\noindent where rms is the spectrum noise as measured in line-free channels, $dv$ is the velocity resolution, and $N$ is the number of channels (e.g. \citealt{2015PASP..127..266M}). Then, we computed the total uncertainty as the quadratic sum of the intrinsic uncertainty plus the calibration uncertainty, so that $\sigma_\mathrm{tot} = \sqrt{\sigma_W^2 + \sigma_\mathrm{cal}^2}$. The calibration uncertainties considered are 14$\%$ for APEX (APEX2, \citealt{2010SPIE.7737E..1JD}) and 5\% for ALMA \citep{2018MNRAS.478.1512B}.

Beam dilution becomes a critical issue, as we work with mean values from single-dish data gathered at different frequencies with different instruments. To compute physically meaningful line ratios, we need to apply a beam-filling factor $F$ to the integrated line intensities. We adopt very conservative estimates of $F\leq$0.2 and $F\leq$0.38 at 230.538 and 345.799 GHz respectively, assuming an emitting area comparable to the ACA beam or smaller.

Hereafter we refer to the CO $J=3\rightarrow2$\, to CO $J=2\rightarrow1$ line ratio as $R_\mathrm{32}$ and to the CO $J=2\rightarrow1$ to $^{13}$CO $J=2\rightarrow1$ as $R_\mathrm{12/13}$. In the ring (component C4) we measure a $R_\mathrm{32} = 0.45\pm0.07$. This value is somewhat low, indicating that the upper level is less populated. On the other hand, we measure a $R_\mathrm{12/13}$ of $25\pm4$ in C4. This value lies halfway between typical ISM values \citep{1992A&ARv...4....1W} and those measured in the outskirts of some evolved massive stars, which show nebular $^{12}$C/$^{13}$C ratios as low as 1--5 \citep{2012ApJ...749L...4L, 2019MNRAS.482.1651B, 2019asrc.confE.101R}, in agreement with theoretical predictions for extremely processed CNO material \citep{2006A&A...447..623M}. But surprisingly, we measure even lower values in the other velocity components. %This indicates that these ratios may be affected by opacity.

% Please add the following required packages to your document preamble:
% \usepackage{booktabs}
\begin{table*}
\caption{Line fitting parameters for each transition: systemic velocity, FWHM, peak temperature, velocity-integrated line intensity and its error. Values have NOT been corrected for the filling factor.}
\label{tab:line-params}
\begin{tabular}{@{}ccccccccccccc@{}}
\toprule
   & \multicolumn{4}{c}{CO $J=2\rightarrow1$}  & \multicolumn{4}{c}{$^{13}$CO$J=2\rightarrow1$} & \multicolumn{4}{c}{CO$J=3\rightarrow2$}   \\ \midrule
   & $v_\mathrm{LSR}$ & FWHM & $T_\mathrm{peak}$ & $W$ ($\sigma W$) & $v_\mathrm{LSR}$ & FWHM & $T_\mathrm{peak}$ & $W$ ($\sigma W$) &   $v_\mathrm{LSR}$ & FWHM & $T_\mathrm{peak}$ & $W$ ($\sigma W$)    \\
   & (km s$^{-1}$) & (km s$^{-1}$) & (K) & (K km s$^{-1}$) &(km s$^{-1}$) & (km s$^{-1}$) & (K) & (K km s$^{-1}$) & (km s$^{-1}$) & (km s$^{-1}$) & (K) & (K km s$^{-1}$) \\ \hline
C1 & 18.7 & 2.2 & 2.19  & 5.01 (0.25)   & 18.6 & 2.2  & 0.24  & 0.54 (0.03)  & 18.6 & 2.2  & 1.31  & 3.00 (0.44)  \\
C2 & 23.8 & 8.9 & 1.25  & 11.9 (0.59)   & 23.5 & 8.9  & 0.08  & 0.71 (0.04)  & 23.7 & 8.9  & 0.56  & 5.27 (0.79)  \\
C3 & 29.6 & 2.3 & 1.57  & 3.78 (0.19)   & 29.6 & 2.3  & 0.11  & 0.26 (0.02)  & 29.6 & 2.3  & 1.05  & 2.53 (0.38)  \\
C4 & 33.7 & 5.7 & 0.79  & 4.79 (0.24)   & 33.7 & 5.7  & 0.03  & 0.19 (0.03)  & 33.7 & 5.7  & 0.67  & 4.09 (0.62)  \\ \bottomrule
\end{tabular}
\end{table*}

To model the excitation of the CO and $^{13}$CO lines, we employed the non-LTE radiative transfer code \textsc{radex} \citep{2007A&A...468..627V}. \textsc{radex} solves the statistical equilibrium equations using the escape probability formulation \citep{1960mes..book.....S} to predict the line intensities and level populations. The code presents two important peculiarities. First, it is entirely agnostic to the source geometry, and hence the predicted line intensities need to be corrected for beam dilution. And second, it works under the assumption of a homogeneous medium, meaning that all transitions sample the same gas volume with the same excitation conditions. These are obvious oversimplifications, yet extremely convenient to provide first-order estimates, especially considering the nature of our data.

We created a grid of models parameterized by the kinetic temperature $T_\mathrm{k}$, the CO column density $N\mathrm{(CO)}$ and the H$_2$ volume density $n(\mathrm{H_2})$. The grid covered $T_\mathrm{k}$ from 10 to 200 K, $N\mathrm{(CO)}$ from 10$^{13}$ to 10$^{19}$ cm$^{-2}$ and $n(\mathrm{H_2})$ from 10$^2$ to 10$^5$ cm$^{-3}$. We used H$_2$ as the only collision partner, with a cosmic background temperature of 2.73 K. We also adopted a representative FWHM of 6 km s$^{-1}$.

\begin{figure*}
 \includegraphics[width=\textwidth]{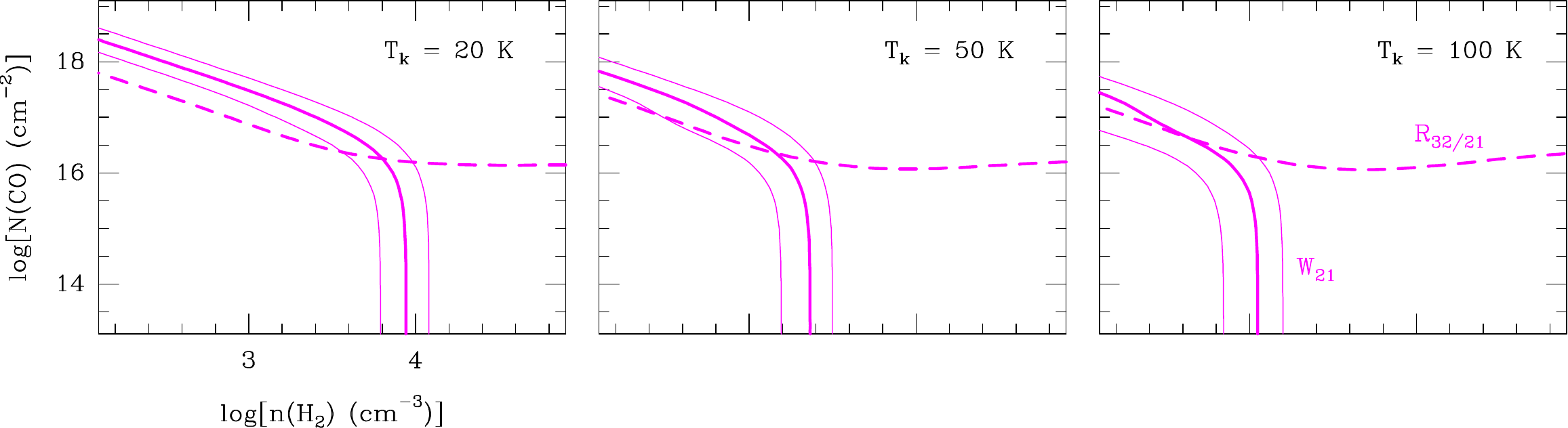}
 \caption{Results of \textsc{radex} fitting for $T_\mathrm{k}$ = 20 K, $T_\mathrm{k}$ = 50 K and $T_\mathrm{k}$ = 100 K. Dashed contours represent the measured CO (3$\rightarrow$2)/(2$\rightarrow$1) ratio; solid contours represent the CO $J=2\rightarrow1$ integrated intensity, with its uncertainties. The intersection of the two sets of contours indicate the range of possible values for $n(\mathrm{H_2})$ and $N\mathrm{(CO)}$.}
 \label{fig:lvg}
\end{figure*}

First, for each $T_\mathrm{k}$ we used $R_\mathrm{32}$ and $W_\mathrm{CO(2-1)}$ to constrain the regions of the parameter space that result in physically meaningful solutions. Sample plots of the fitting are shown in Fig. \ref{fig:lvg}.  We find that $N(\mathrm{CO})$ is well determined, being relatively insensitive to temperature and density and taking values of a few 10$^{16}$ cm$^{-2}$. On the other hand, $n(\mathrm{H_2})$ is only loosely constrained, with valid solutions in the range 10$^3$--10$^4$ cm$^{-3}$. Regarding the temperature, though, solutions for $T_\mathrm{k}>120$ K are unlikely: the required volume densities are comparable to those of diffuse clouds ($\sim$ 100 cm$^{-3}$), falling well below the critical density of the $J=2\rightarrow1$ transition. Thus we favour solutions with $T_\mathrm{k}<100$ K.

\begin{table}
\begin{tabular}{@{}cccccc@{}}
\toprule
          & $T_\mathrm{k}$  & $N\mathrm{(CO)}$ & $n(\mathrm{H_2})$ & $\tau_{21}$ & $\tau_{32}$ \\ 
          & [K]   & [10$^{16}$ cm$^{-2}$]  & [10$^{3}$ cm$^{-3}$]   &       &       \\\midrule
          & 20  & 1.8   & 5.6   & 0.8   & 0.6   \\
Ring (C4) & 50  & 2.4   & 1.3   & 1.2   & 0.8   \\
          & 100 & 4.2   & 1     & 2.4   & 1.2   \\
C1        & 30  & 0.24  & 10    & 0.1   & 0.2   \\
C2        & 30  & 0.75  & 4.2   & 0.2   & 0.2   \\
C3        & 30  & 0.18  & 10    & 0.1   & 0.1  \\ \bottomrule
\end{tabular}
\caption{LVG results for the four components under analysis. From left to right, kinetic temperature, CO column density, H$_2$ volume density and optical depth. Values for components C1 to C3 are computed for a reference temperature of 30 K.}
\label{tab:lvg}
\end{table}

\begin{figure}
 \includegraphics[width=\columnwidth]{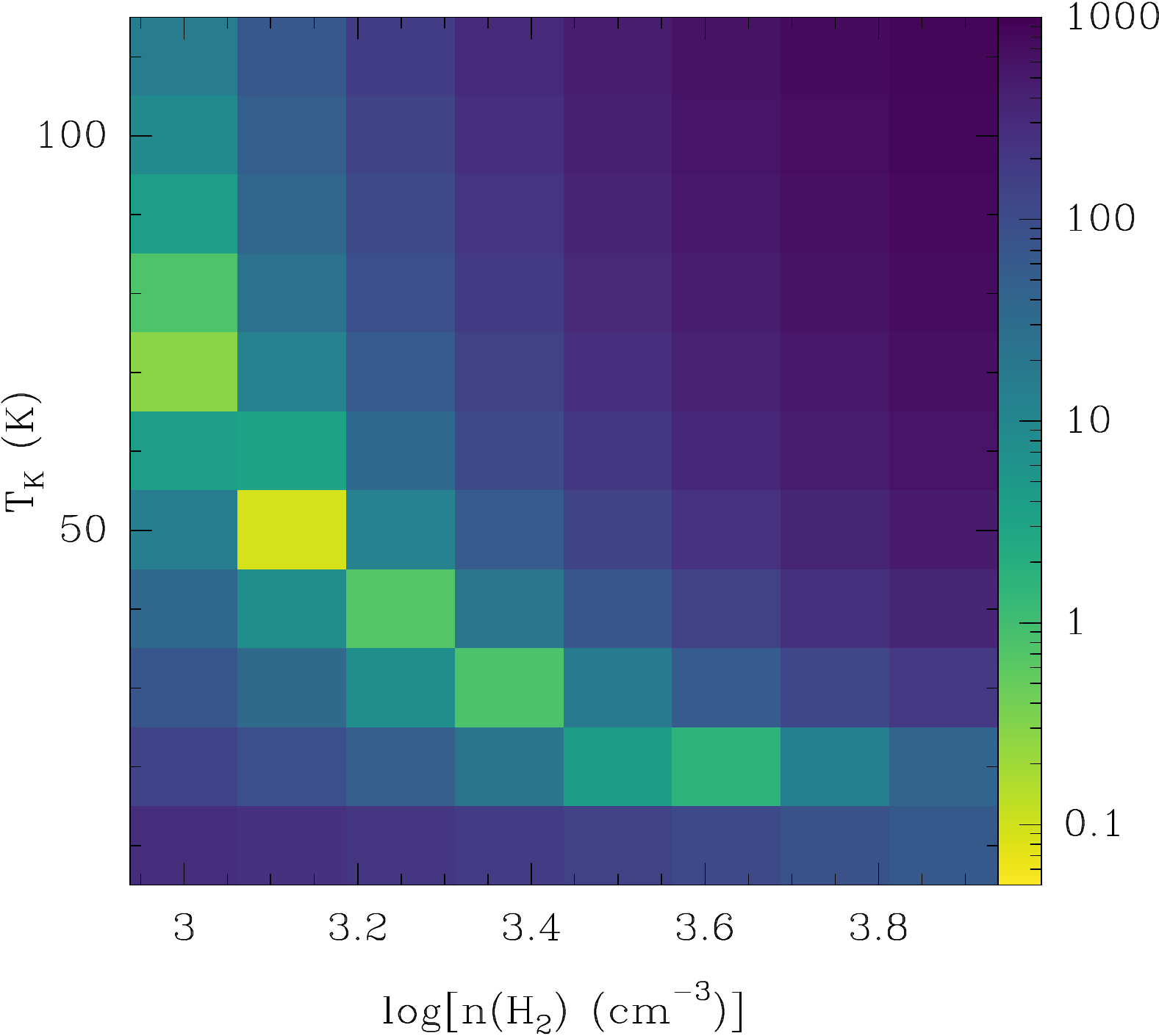}
 \caption{$\chi^2$ surface of the \textsc{radex} models, in logarithmic color scale, as a function of $T_\mathrm{k}$ and $n(\mathrm{H_2})$. The models in this plot correspond to the best-fitting $N(\mathrm{CO}) = 2.4\times10^{16}$ cm$^{-2}$.  }
 \label{fig:chi}
\end{figure}

Once we constrained the parameter space, we followed a reduced $\chi^2$ minimisation technique to find the model that better reproduces the observed line intensities, such that

\begin{equation}
\chi^2 =\sum_{i=1}^{2} \left(\frac{W_i^\mathrm{rad} - W_i^\mathrm{obs}/F_i}{\Delta W_i^\mathrm{obs}} \right)^2
\end{equation}

\noindent where $W_i$ is the integrated intensity of the $(i+1)\rightarrow i$ transition and $F_i$ is the corresponding filling factor. Fig. \ref{fig:chi} shows the fit $\chi^2$ surface as a function of $T_\mathrm{k}$ and $n(\mathrm{H_2})$. We find that the model that best fits the observations is that with $T_\mathrm{k}$ = 50 K, $N(\mathrm{CO})$ = 2.4$\times10^{16}$ cm$^{-2}$ and $n\mathrm{(H_2)}$ = 1.3$\times10^3$ cm$^{-3}$. The solution points to a moderately thick opacity in the emitting medium, with optical depths of $\tau_{21}=1.2$ and $\tau_{32}=0.8$ for this temperature range. As a cross check, we used the derived parameters to predict the $^{13}$CO $J=2\rightarrow1$ integrated intensity, for a range of $^{13}$CO column densities. The best match is obtained for $N(^{13}\mathrm{CO}) = 7.5\times10^{14}$ cm$^{-2}$, which yields an isotopic ratio of $\sim$ 30. This value agrees with the $R_\mathrm{12/13}$ previously discussed.

Finally, we repeated the same procedure to study the excitation conditions of the other components, C1 to C3. In this case, as the emission is widespread --in fact, much larger than the beam-- we did not apply a correction for beam dilution. We obtained column densities below 10$^{16}$ cm$^{-2}$, but we were unable to properly constrain the kinetic temperature of the gas. We found many possible solutions, starting from the typical temperatures of a cold cloud, around $\sim$ 10 K. All these solutions occurred roughly at constant pressure ($n\times T$), which suggest an equilibrium situation. In addition, the lines were always more optically thin ($\tau<1$), in contrast with the gas of the ring, which is significantly more opaque. These differences suggest that component C4 is indeed subject to slightly different excitation conditions. LVG results of the four components are compiled in Table \ref{tab:lvg}.

\subsubsection{Estimate of the ring mass}

We can use the average CO column density derived with \textsc{radex} to provide a crude estimate of the mass of the ring. The total mass of molecular gas is given by

\begin{equation}
    M = 2 \mathrm{m}_\mathrm{H}\,f_\mathrm{He}\, N\mathrm{(CO)}\, X_\mathrm{CO}^{-1}\,\Omega_\mathrm{sou}
\end{equation}

\noindent where $\mathrm{m}_\mathrm{H}$ is the mass of the hydrogen atom, $f_\mathrm{He}$ is the He correction factor, $X_\mathrm{CO}$ is the relative [CO/H$_2$] abundance and $\Omega_\mathrm{sou}$ is the solid angle subtended by the source. Adopting a cosmic [CO/H$_2$] abundance of $X_\mathrm{CO} = 10^{-4}$  and considering that the ring subtends an area of about $\sim0.63$ pc$^2$ at $d=6$ kpc, we obtain a total mass of 2.7 $\pm$ 0.9 $\mathrm{M}_{\sun}$, a result comparable with molecular gas masses found in other LBVN: in $\eta$ Car, the equatorial torus contains 1--5 $\mathrm{M}_{\sun}$ of molecular material \citep{2018MNRAS.474.4988S}, and the molecular ring of MN101 has a mass of 0.6$\pm$0.1  $\mathrm{M}_{\sun}$ \citep{2019MNRAS.482.1651B}. %Our estimate is actually very similar to the lower limit proposed by \cite{2002AJ....124.2920N} of $\sim$ 2.8 $\mathrm{M}_{\sun}$.

It is worth to highlight that this estimate is strongly affected by the determination of the $X_\mathrm{CO}$ factor and its associated uncertainties. Many LBVN are deficient in C and O, and consequently underabundant in CO (e.g. $\eta$ Car, \citealt{2017ApJ...842...79M}). Should this be the case for AG Car as well, the mass of the ring could be substantially higher.

\subsection{Origin of the molecular ring}
\label{sec:origin}

\begin{figure}
 \includegraphics[width=\columnwidth]{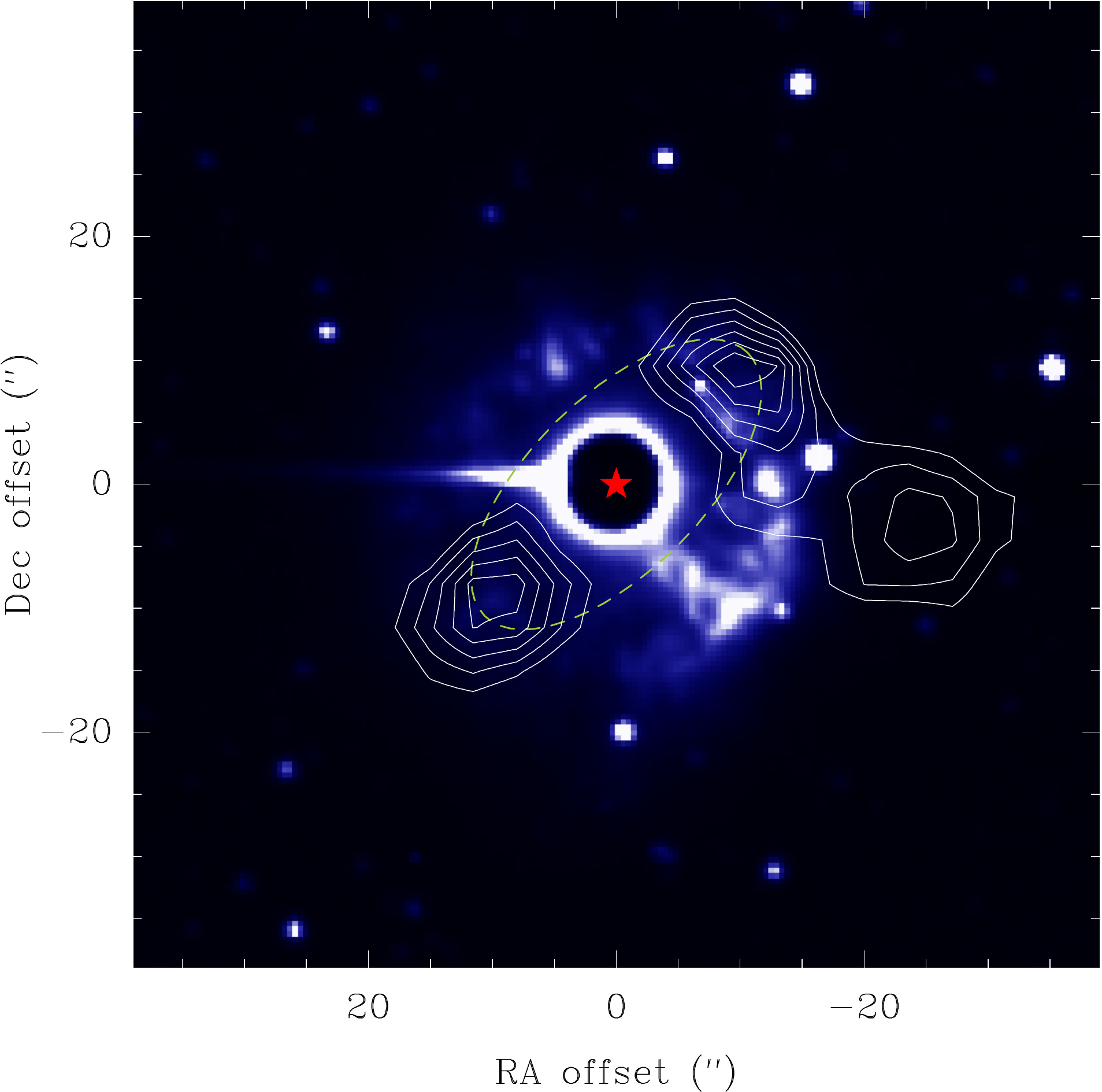}
 \caption{Relative orientation of the optical jet and the molecular ring. EFOSC1 coronographic continuum image of AG Car and its nebula in color scale \citep{2015A&A...578A.108V}. ACA+TP CO $J=2\rightarrow1$ intensity map superimposed as contours. Contours start at 50\% of the peak intensity, showing only the ring blobs, for the sake of clarity. The dashed ellipse is displayed to highlight the extent of the structure. The red marker indicates the position of the star. The bright spike to the east is an imaging artefact.}
 \label{fig:jet}
\end{figure}

As shown in previous sections, the ring surrounding AG Car is composed of moderately dense, warm gas slowly expanding into the ISM. The physical connection between the molecular gas and the star, as inferred from morpho-kinematic features and excitation conditions, seems evident. The ring is seen nearly edge-on, with an inclination of $\sim$70\degr. This is in excellent agreement with the viewing geometry proposed by \cite{2006ApJ...638L..33G, 2009ApJ...705L..25G}, who measured a remarkably high rotation velocity for the star ($v\sin{i}=220$ km s$^{-1}$) and concluded that we necessarily see AG Car from the equator. Considering this, we can safely assume that the ring lies in --or at least close to-- the equatorial plane of the star. Such an assumption is further supported by the orientation of optical continuum jet found by \cite{1992ApJ...398..621N}, which should be relatively aligned with the rotation axis of AG Car. Fig.\ref{fig:jet} shows the EFOSC (ESO Faint Object Spectrograph and Camera) optical continuum image of AG Car, with the CO contours superimposed. The ring and the jet appear perfectly perpendicular in projection, with P.A. of 135 and 225\degr\ respectively. However, an equatorial ring may be the result of very different physical processes. Below we discuss possible formation scenarios and their implications.

\subsubsection{Swept-up pre-existing material}

One possibility is that CO traces a remnant of the parent molecular cloud in which the star formed, that has been compressed by the action of the stellar wind. However, this interpretation involves some problems, as massive stars like AG Car sustain very fast winds of $\sim$ 1000 km s$^{-1}$ in the main sequence. For several Myrs, these winds sweep up the stellar neighbourhood, eroding and even completely destroying the natal clouds and carving humongous cavities in the ISM. These so-called wind-blown bubbles have scales much larger than the size of ring, even spanning several tens of pc \citep{1996A&A...305..229G}. Should the CO arise from compressed ISM material, one would expect to find it farther from the star, toward the edges of a wind-blown bubble. \textit{Herschel} images revealed that AG Car is in fact immersed within a cavity about 5\arcmin ($\sim9$ pc) in diameter, where material has been mostly evacuated (see fig.3 in VN15).

Even so, as \cite{2002AJ....124.2920N} pointed out, aspherical winds due to stellar rotation could result in slower velocities near the equatorial plane of the star, thus 'protecting' ISM material at low latitudes. Yet, the long-term survival of a compact structure so close to the star is mainly determined by the wind density: a fast but not very dense stellar wind would struggle to evacuate the densest parts of a molecular cloud. The dynamical age of the ring, however, favours a more recent, post-main sequence origin, not more than $\sim$10$^5$ years ago.

\subsubsection{A binary interaction remnant}

As a second formation hypothesis, one may invoke a binary scenario. The possible binary nature of AG Car has been long discussed in literature ever since the discovery of its bipolar nebula, and particularly, its helical jet, which strongly suggests some kind of precession \citep{1989ApJ...341L..83P, 1992ApJ...398..621N}.

In recent times, the traditional view of LBV stars in the frame of single-star evolution has been challenged. The apparent isolation of most LBVs from OB associations, their phenomenological heterogeneity as a class, and their role as potential SN progenitors, have led some authors to propose an alternative evolutionary paradigm, in which LBVs likely arise from binary interaction. Under this novel framework, two possible formation pathways are considered: either LBVs are \textit{rejuvenated} mass gainers in close binary systems, that end up kicked out by their companion's SN; or they are produced following a binary merger event \citep{2014ApJ...796..121J, 2015MNRAS.447..598S, 2016MNRAS.461.3353S, 2017MNRAS.472..591A}.

Assuming that AG Car is part of a close interacting binary, a non-conservative Roche Lobe overflow (RLOF) episode would result in a mass-leakage through the outer Lagrange point of the system (L2). This leakage would effectively create a ring-like structure, slowly expanding outwards. Besides, the gainer not only accretes mass in the process, but also angular momentum, speeding up its rotation, which would be consistent with the enhanced rotation of AG Car. This kind of non conservative mass-transfer interaction has been previously suggested as the driving mechanism in other non-LBV sources that exhibit equatorial ring-like structures, such as the less massive binary RY Scuti \citep{2011MNRAS.418.1959S} or SBW1, a very luminous, 18--25 $\mathrm{M}_{\sun}$ BSG \citep{2007AJ....134..846S}. Still, any assumptions on the binarity of AG Car are highly speculative and must be regarded with caution. So far, the search for a companion has been unfruitful, yielding negative results even at X-ray wavelengths \citep{2012A&A...538A..47N}. However, spectroscopic signatures of potential companions may be hidden by the rotationally broadened lines of AG Car. In any case, binarity does not seem to be an extremely common feature among the LBV family, with only a handful of them being present-day binaries, such as $\eta$ Car \citep{2000ApJ...528L.101D} and HR Car \citep{2016A&A...593A..90B}.

On the other hand, the binary merger hypothesis allows us to circumvent the lack of observational evidence for a companion of AG Car. The merger mechanism described by \cite{2014ApJ...796..121J} depicts a massive binary in which the primary, expanding as it evolves beyond the main sequence, fills its Roche lobe transferring mass onto the secondary. For certain mass ratios, the system enters a contact phase due to expansion of the secondary's envelope. Eventually, the system destabilizes and a merger occurs. This mechanism could also lead to the formation of rings, as equatorial mass outflows are expected during the merging process as a consequence of the common envelope rotation. The merger idea, which was already introduced two decades ago for some B[e] stars \citep{1998ASSL..233..235L, 2000AJ....119.1352P}, has recently become one of the preferred explanations for $\eta$ Car's Great Eruption in the 1840s. The outburst would have been triggered by a merger within a hierarchical triple system, leading to the current LBV + O/WN binary \citep{2016MNRAS.456.3401P, 2018MNRAS.480.1466S, 2020A&A...640A..16T} and giving rise to an equatorial torus. While it is difficult to infer whether AG Car has undergone a similar process, we note that many of its features (bipolarity, enhanced rotation, He and N abundances) are predicted for the BSG/LBV products of massive binary mergers, making this hypothesis a fascinating possibility worth to be further explored.

In the general picture, the portrait of LBVs as products of binary interaction seems able to explain many of their peculiarities. The formation of the slowly expanding triple ring system in SN1987A has been equally linked to a possible binary merger \citep{2007Sci...315.1103M}, promoting binarity and rotation as two key ingredients of the sometimes elusive connection between LBVs and supernovae.

\subsubsection{An equatorially enhanced outflow}

Finally, another possible explanation sticking to the traditional single-star scenario is that the molecular ring in AG Car originated from an equatorially enhanced mass-loss episode. This kind of non-spherical mass-loss seems to be a common phenomenon in LBV stars, as many of them are surrounded by dusty or gaseous equatorial tori, such as $\eta$ Car \citep{1999Natur.402..502M,2018MNRAS.474.4988S}, HD168625 \citep{2003ApJ...598.1255O}, and MN101 \citep{2019MNRAS.482.1651B}. However, little is known about the mechanisms behind the formation of disks or ring-like structures around single massive stars. 

Stellar rotation is thought to play a crucial role in the shaping of stellar winds throughout the different evolutionary stages of massive stars. In this context, we may explain the formation of a gaseous ring in AG Car by invoking a rotationally-induced bistability mechanism. The bistability jump would produce increased mass fluxes and slower winds at low latitudes near the stellar equator, as a direct outcome of the decrease in the effective gravity $g_\mathrm{eff}$ and the subsequent gravity darkening \citep{1991A&A...244L...5L, 2000A&A...359..695P}. Similarly, at higher latitudes, the winds would be faster but less dense. 

By virtue of this mechanism, aspherical winds and equatorially enhanced mass-loss are strongly favoured in fast-rotating stars with high radiation pressures \citep{2001A&A...372L...9M}.  AG Car, which rotates at a significant fraction of its break-up velocity at least during its hottest phase (up to 86\%, \citealt{2006ApJ...638L..33G}), satisfies these two conditions. \cite{1994ApJ...428..292L} found two independent pieces of evidence that support this scenario: (1) a significant --and variable-- degree of polarization in AG Car, interpreted as an equatorial density enhancement of the stellar wind, and (2), a two-component wind structure: a slow and dense wind, traced by recombination lines, in coexistence with a faster and less dense component, seen in ultraviolet absorption lines.

Another important issue to address is the survival of molecular gas in the proximity of AG Car. For molecules to form out of the stellar wind, a certain degree of shielding against the strong FUV radiation must be provided. In this regard, AG Car may somehow be an analog of B[e] supergiants, where slowly expanding, rotating disks or rings of neutral material have been observed.  The formation mechanism of such structures has been tentatively linked to nearly-critical stellar rotation as well \citep{1986A&A...163..119Z, 1998MNRAS.300..170O, 2005A&A...437..929C, 2006A&A...456..151K, 2010A&A...517A..30K}. The equatorial winds of B[e]SGs provide the adequate conditions, in terms of temperature and density, for the formation and survival of significant amounts of dust and molecules \citep{2010MNRAS.408L...6L}, but B[e]SG disks are much smaller, spanning only a few hundreds of AU. In the case of AG Car, the expansion of the wind over large scales would involve an important decrease in density. Perhaps, an inhomogeneous stellar wind --which is a common feature in many hot, massive stars, see e.g. \cite{2004RMxAA..40...53C}-- would make possible the formation of molecules within denser clumps, that would provide the necessary protection against ionizing photons. Interestingly, modelling of the nebular dust content by \cite{2000A&A...356..501V} predicts a population of very large grains (with sizes up to 40 $\mu$m), which are typically found in circumstellar disks. If such a disk exists in AG Car, it would effectively shield the gas, favouring the formation of molecules in its cold outskirts.

Our estimate of the dynamical age of the ring is slightly high compared to the characteristic timescales of the LBV phase, a few 10$^3$--10$^4$ years. Thus the structure may have formed in a pre-LBV stage. Along this line of reasoning, some authors have proposed that the main nebula of AG Car was expelled before reaching the LBV phase. \cite{1997MNRAS.290..265S}, \cite{2000A&A...356..501V} and \cite{2001ApJ...551..764L} found nebular abundances of N and O different from the values expected for CNO-enriched material, suggesting that the nebula is composed of 'mildly processed' matter, still far from CNO-equilibrium and thus consistent with older ejecta from a previous BSG or RSG phase. These findings, although not entirely conclusive, might also explain the relatively high $R_\mathrm{12/13}$ that we measure, more compatible with a moderate degree of processing.

Likewise, the low expansion velocity of the ring is another interesting issue. If we work under the hypothesis that the ejection took place in a pre-LBV stage, a $v_\mathrm{exp}$ of 3.5 km s$^{-1}$ is in agreement with the typical velocities of RSG winds, of 5--10 km s$^{-1}$. The possible occurrence of an RSG phase in AG Car, however, needs to be investigated, in view of the apparent lack of very luminous RSGs in the HR diagram (the so-called 'red-supergiant problem', \citealt{2012IAUS..279..419W}). On the other hand, BSG winds are faster and more compatible with the expansion velocity of $\sim$70 km s $^{-1}$ measured in the main nebula. This probably indicates that the ring originated in a separate episode of mass-loss, which may have been relatively steady, in contrast with the main nebula, related to a more eruptive event. Never the less, the measured expansion velocity only reflects the current situation, but the wind may have been faster in the past, resulting in a younger structure as discussed in Sect. \ref{sec:morph-kin}. In such case, depending on the age, we may obtain time-scales consistent with an ejection event during the LBV phase, that would require further explanations for the observed chemistry.

\subsection{A closer view to CO kinematics}
\label{sec:kin}

We have built a simple model to better understand the kinematics of the ring, and also to cross-check the validity of the \textsc{radex} fit. To do so, we used \textsc{lime} \citep{2010A&A...523A..25B}, a 3D non-LTE code that solves radiative transfer by simulating ballistic photon propagation through unstructured Delaunay grids, calculating level populations and predicting the resulting spectra for a given transition, according to the LAMDA database collisional coefficients \citep{2005A&A...432..369S}.

\begin{figure*}
 \includegraphics[width=\textwidth]{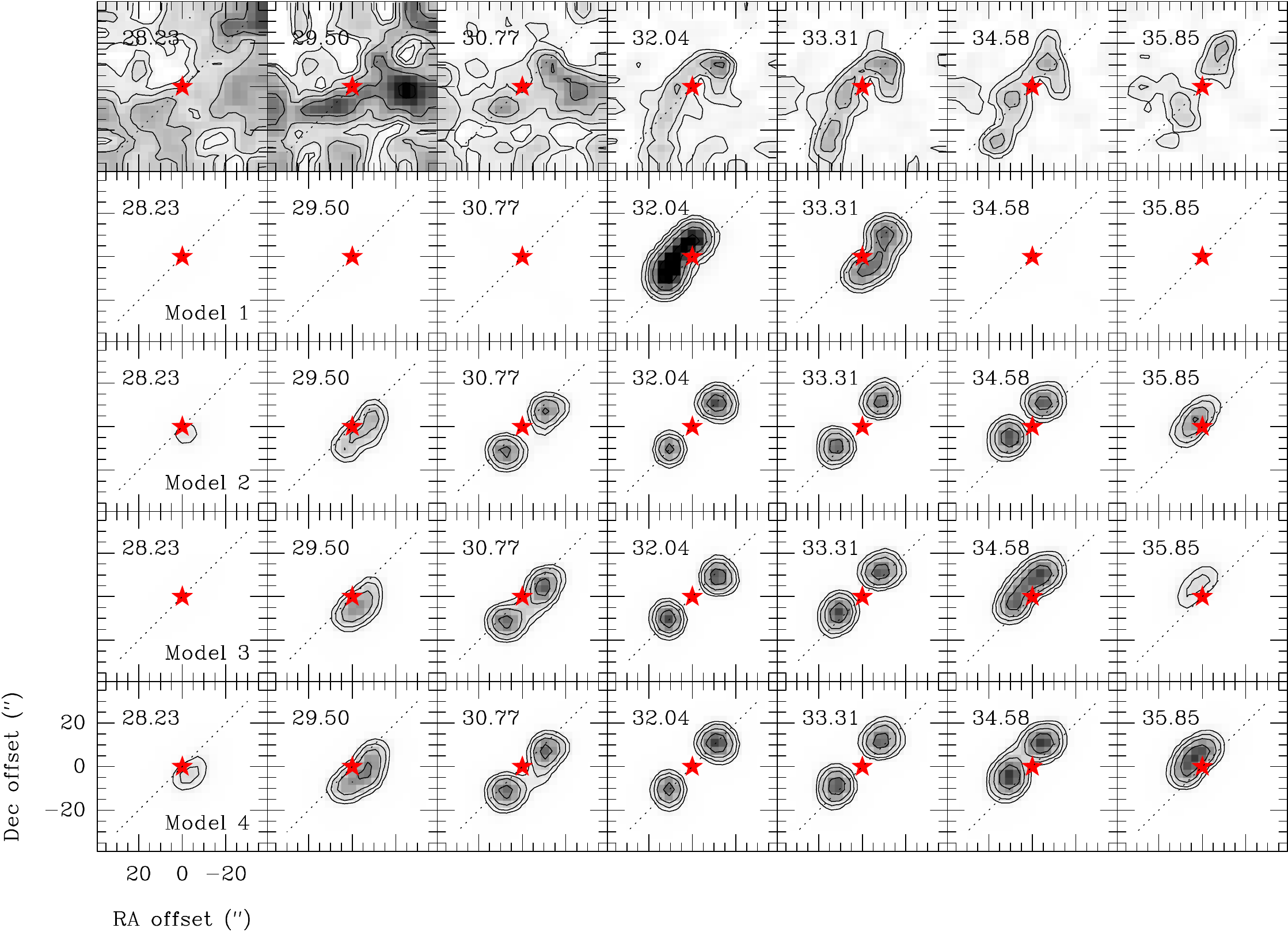}
 \caption{Comparisons of the observed CO $J=2\rightarrow1$ channel maps (top row) with different kinematic models: a rotating ring, a purely expanding ring, a expanding ring with rotation and an expanding ring with macroturbulence. All the maps share the same intensity scale. Contours at 0.5, 1, 2, 4 and 8 K. The red marker indicates the position of the star and the dashed line represents the direction of the major axis of the structure. The LSR velocity is displayed in the top-left corner of each panel. }
 \label{fig:mdl-chan}
\end{figure*}

Due to the modest spatial and spectral resolution of our data, we made several assumptions to keep the model as simple as possible; we supposed an axisymmetric ring located at 6 kpc and described by power-law density and temperature profiles of the form $r^{-2}$. To avoid overinterpreting our maps, instead of fitting the observed data, we qualitatively compared the overall morpho-kinematic features in the structure with those obtained with different kinematic models. Therefore, we adopted the density and temperature outputs of \textsc{radex} as reference values for the power-law profiles. Regarding the geometrical parameters, we adopted an inner radius of 0.3 pc, a half opening angle of 10\degr, an inclination of 70\degr and a position angle of 135\degr east of north.

We created four models to explore different possibilities for the velocity structure of the gas, namely: a purely rotating ring with differential rotation (model 1); a radially expanding ring with an outflow velocity of 3.5 km s$^{-1}$ (model 2); an expanding ring with a differential rotation component (model 3); and an expanding ring with macro-turbulence, i.e. with random departures from the expansion velocity law, of up to 0.25 km s$^{-1}$ (model 4). Model parameters are compiled in Tab. \ref{tab:mdl-params}. For each of the models, \textsc{lime} produced a synthetic CO $J=2\rightarrow1$ cube, which was later convolved to the ACA beam and re-gridded to allow for a direct comparison with the original data.

\begin{table}
\begin{tabular}{@{}lll@{}}
\toprule
Model & Velocity law              & Parameters                        \\ \midrule
1     & Rotation                  & $v_\phi = 1$ km s$^{-1}$                   \\
2     & Expansion                 & $v_\mathrm{exp} = 3.5$ km s$^{-1}$                         \\
3     & Expansion + rotation   & $v_\phi = 1$ km s$^{-1}$, $v_\mathrm{exp} = 3.5$ km s$^{-1}$ \\
4     & Expansion + turbulence & $v_\mathrm{exp} = 3.5$ km s$^{-1}$, $v_\mathrm{turb} = 0.25$ km s$^{-1}$     \\ \bottomrule
\end{tabular}
\caption{Model parameterization. Rotation velocities, $v_\phi$, refer to the inner radius of the ring.}
\label{tab:mdl-params}
\end{table}

Fig. \ref{fig:mdl-chan} compares the four models as channel maps in the velocity range of interest. Not surprisingly, the rotating ring (model 1) fails to reproduce the shape and velocity extent of the ring, with all the emission significantly concentrated around the systemic velocity. The other three models are able to reproduce both the extent of the ring and the overall clockwise pattern satisfactorily. Still, we note that the arc-like features that connect the lobes at intermediate velocities (from 29.5 to 30.77 and 34.58 to 35.85 km s$^{-1}$) can only be reproduced by models 3 and 4, i.e. if we add rotation or turbulence to the expanding ring. None of the models reproduces the asymmetric spatial distribution between 32 and 33.3 km s$^{-1}$ correctly. Ring inhomogeneities or clumpiness, not considered here, may account for the observed differences.

\begin{figure*}
 \includegraphics[width=\textwidth]{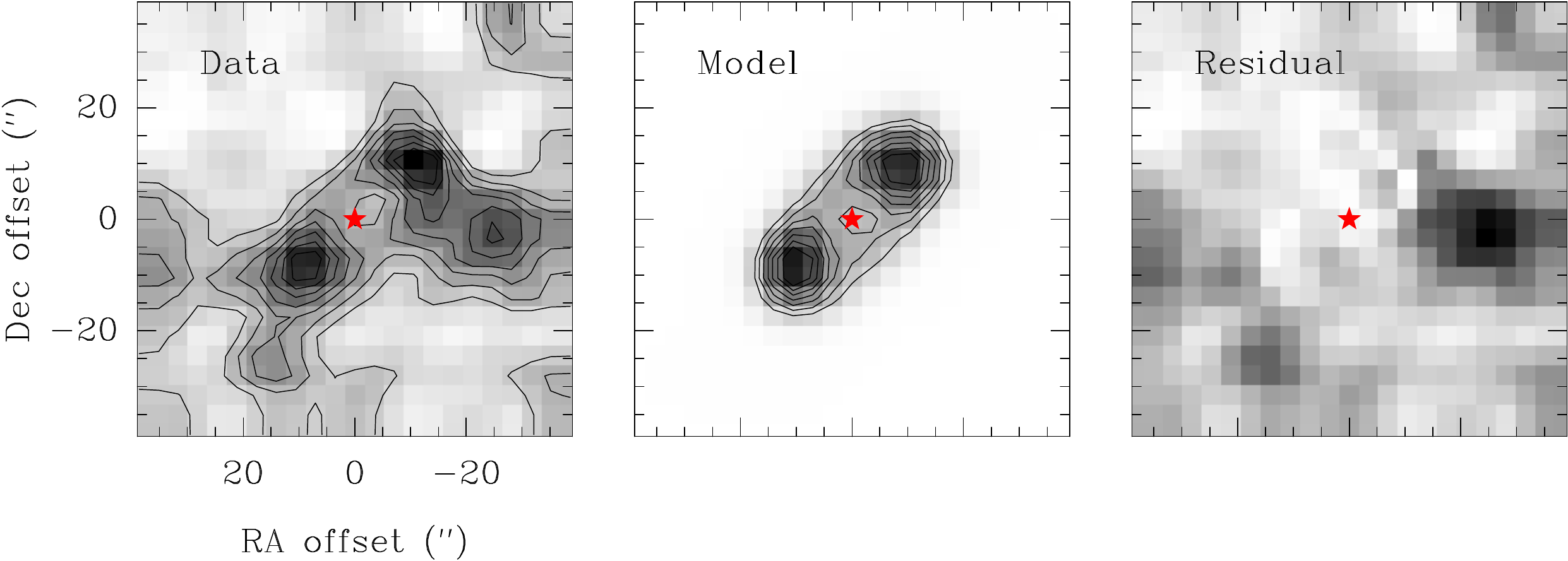}
 \caption{Comparison of the observed CO $J=2\rightarrow1$ intensity map with that of model 3. The residual map is shown for reference. The red marker indicates the position of the star. }
 \label{fig:mdl}
\end{figure*}

Models 3 and 4 also reflect the overall morphology of the integrated intensity map from 29 to 36 km s$^{-1}$ (e.g. Fig. \ref{fig:mdl} for model 3). Indeed, we found the integrated fluxes to be consistent within a $\sim$10$\%$, thus confirming that the average values for density and temperature obtained with \textsc{radex} are able to reproduce the observed intensities.

While the presence of turbulent motions in the wind of an unstable star is somewhat expected, the physical feasibility of the expanding-rotating ring model deserves a comment. In principle, any material expelled from the stellar surface, whatever the mechanism for triggering the ejection, will tend to move radially in the long-term. In the case of a fast-rotator like AG Car, the ejecta could be initially lifted with a non-negligible rotational component to preserve angular momentum. However, this component will progressively fade as the ring dissipates radially and interacts with the surrounding medium. Moreover, to keep a stable rotation, the material needs to be either gravitationally or magnetically bound to the star. These mechanisms, though, operate only at short distances, but the molecular ring of AG Car is $<10^5$ years old and far away from the star. Consequently, if the gas actually rotates as indicated by model 3, there should be another mechanism at work responsible for that rotation. Further observations at higher velocity resolution will allow for a better understanding of the gas dynamics.

\section{Mass loss history of AG Car}

The different evolutionary stages of AG Car have left a footprint in its environment, shaping its circumstellar medium (CSM) in complex manners due to variable stellar winds and episodic mass eruptions. The detection of a molecular ring surrounding the star provides additional information about its recent mass-loss history. It is therefore important to understand how this molecular gas relates to the other components in the CSM. 

\begin{figure*}
 \includegraphics[width=\textwidth]{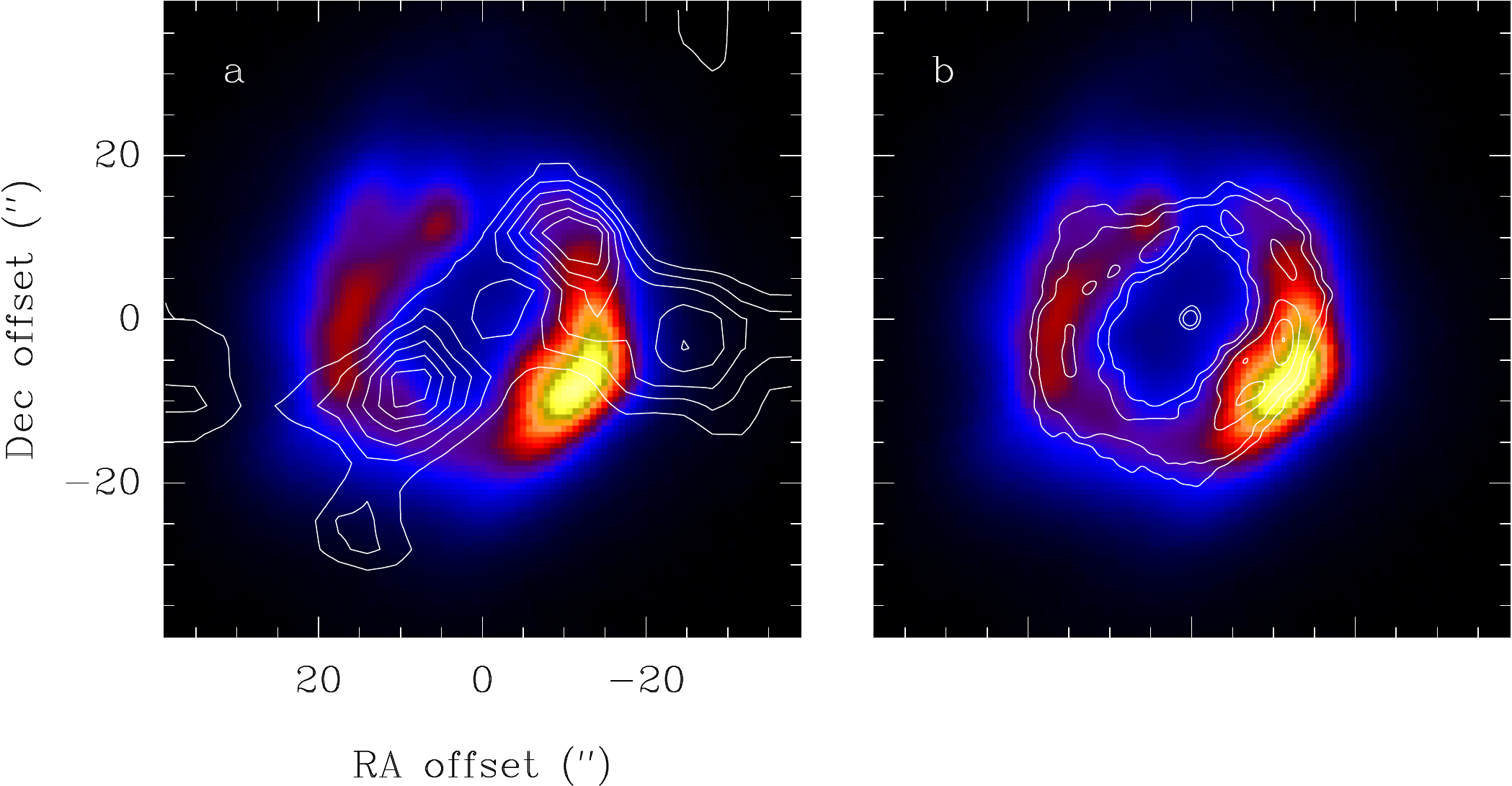}
 \caption{Multi-wavelength comparison of the nebular material around AG Car. (a) \textit{Herschel} PACS 70 $\mu$m image in color scale with the ACA CO $J=2\rightarrow1$ contours superimposed. The levels are the same as in Fig. \ref{fig:aca}.(b) The same as (a) but with the ATCA radio continuum contours at 5.5 GHz \citep{2002MNRAS.330...63D}}
 \label{fig:ircomp}
\end{figure*}

Broadly speaking, the CSM around AG Car can be described in terms of two independent nebular structures: a detached dusty shell, bright in the infrared; and an ionised nebula, clearly visible in optical and radio continuum images. These structures present strong morpho-kinematic hints of bipolarity. Across the NE-SW direction, two symmetric bright clumps at P.A. $\sim$35 and $\sim$225\degr are evident. In addition, the measured expansion velocity of the shell increases notably (by about 20 km s$^{-1}$) along this direction \citep{1992ApJ...398..621N}. Both the orientation of the clumps and the velocity increase are signposts of a bipolar mass outflow, in which case the bright clumps would correspond to gas density enhancements in the polar 'caps' of the structure, where more material is being accumulated. Fig. \ref{fig:ircomp} shows the location of the CO gas with respect to the dust and the ionised structures. Although an accurate spatial comparison is not possible given the limited resolution of the molecular data, we note that, while the dust and the ionised gas are mostly co-spatial, the molecular gas seems to be complementary: the most intense CO clumps are found along the SE-NW axis, i.e. the direction of the slight elongation of the nebula \citep{2000A&A...356..501V}. 

The molecular ring somewhat constrains the overall geometry of the CSM: an inhomogeneous, nearly spherical expanding shell, disrupted by a slightly faster bipolar outflow and enclosed by an equatorial density enhancement. What is clear, though, is that the equatorial density enhancement should be older than the shell, since the two structures have comparable scales but very different expansion velocities. This geometry, coherent from an evolutionary perspective, can be explained attending to the stardard wind-interaction model, which has been proposed to explain the morphologies of other LBVN with aspherical symmetries (e.g. \citealt{1995ApJ...441L..77F}). In this particular scenario, we can imagine AG Car suffering an equatorial mass-loss enhancement, in the form of a dense, steady wind, some $\sim10^5$ years ago. This wind would slowly expand into the ISM, sweeping any remaining ambient material. At some point, the wind would get colder and form molecules. A few $\sim10^4$ years later, a faster isotropic outflow may take place. This new wind would eventually reach the slower, previous one, interacting with a torus-like density distribution. The result of such interaction is the formation of a bipolar nebula. The final shape --i.e. the degree of bipolarity-- is modulated by several factors, including the rotation velocity of the star \citep{2001A&A...372L...9M}, and may range from an ellipsoid to a peanut-like structure. The most extreme case of such interaction is, of course, $\eta$ Car and the Homunculus Nebula. In this sense, VN15 suggest that the ejection of the shell took place in a period of slow rotation, which is consistent with the moderate degree of bipolarity observed.

VN15 estimated a total mass of ionised gas of $6.6\pm1.9 \,\mathrm{M}_{\sun}$, from the H$\alpha$ and radio continuum fluxes. Similarly, they derived a dust mass of $0.20\pm0.05 \,\mathrm{M}_{\sun}$ from the SED modelling. Using our continuum data and following their approach, we obtain a nebular ionised mass of 3.9$\pm1.9\, \mathrm{M}_{\sun}$, smaller but still in agreement within uncertainties with VN15's estimate. However, as noted in Sect. \ref{sec:results:continuum}, we may be missing some flux with ACA. The total combined mass of ionised and molecular gas thus adds up to $9.3\pm2.8\,\mathrm{M}_{\sun}$ (or $5.9\pm2.8 \,\mathrm{M}_{\sun}$ if we take our estimate). This agrees again, within uncertainties, with the value proposed by VN15 ($15\pm4.5\,\mathrm{M}_{\sun}$), who adopted a gas-to-dust ratio of $\sim40$ to estimate the contribution of neutral gas. This means that, assuming a stellar origin, the molecular gas around AG Car traces at least a $\sim30\%$ of the mass lost by the star.

The gas and dust masses can be translated into mass-loss rates by estimating the kinematic duration of the events that produced the observed structures. This can be done by computing the temporal difference between the inner and outer shell radius for a constant expansion velocity. Following this method, VN15 derived a kinematic duration of $1.1\times10^4$ years for the mass-loss episode that created the infrared nebula, taking an inner and outer radii of 0.4 and 1.2 pc respectively. It is a useful exercise to do the same and provide a crude estimate of duration of the molecular outflow. However, we warn that the reliability of this method is limited, in the sense that the emitting region may be smaller than the beam; therefore, we decided to conservatively adopt the size of the ACA beam, of about 7\arcsec, as the reference 'width' of the ring. In addition, we work under the assumption that the expansion velocity does not change and there is no velocity dispersion across the ring, which is probably unrealistic. For $v_\mathrm{exp}=3.5$ km s$^{-1}$, we obtain an approximate duration of $5.6\times10^4$ years. This is just a loose upper limit, but still, it is interesting to realise that it is about five times larger that the duration of VN15 for the isotropic mass-loss episode. This probably suggests that the equatorial enhancement was more steady and less eruptive, which is, again, in line with the proposed formation scenario.

The average mass-loss rate that produced the molecular ring is thus $\dot M = (4.8\pm1.6)\times10^{-5}\,\mathrm{M}_{\sun}$ yr$^{-1}$, but it could be substantially higher if the structure is more compact than the ACA beam. We can compare this value with the current mass-loss rate. Under the assumption that the flux density of the point-like source measured at 225 GHz is due to stellar wind, we can approximate the current mass-loss rate of AG Car using the empirical formula by \cite{1975A&A....39....1P} for expanding envelopes around hot stars, in which we assume full ionization and standard cosmic abundances, so that

\begin{equation}
\dot M = 6.7\times10^{-4} v_\infty F_\nu^{3/4} d^{3/2} (\nu \times g_\mathrm{ff})^{-0.5}
\end{equation}

\noindent where $v_\infty$ is the terminal wind velocity in km s$^{-1}$, $d$ is the distance in kpc, $F_\nu$ is the measured flux in mJy and $\nu$ is the central frequency in Hz. $g_\mathrm{ff}$ is the free-free Gaunt factor, which we approximate as $g_\mathrm{ff} = 9.77(1+0.13\log{\frac{T^{3/2}}{\nu}})$, where $T$ is the wind plasma temperature \citep{1991ApJ...377..629L}. Providing a single value for the terminal wind velocity of AG Car is tricky. Long-term spectroscopic monitoring by \cite{2001A&A...375...54S} between 1989 and 1999 revealed abrupt changes from 225 to 30 km s$^{-1}$. The wind also changes drastically between consecutive visual minima, with values of 300 km s$^{-1}$ between 1985-1990 and 105 km s$^{1}$ in 2000-2001. Therefore, since we are now on the way to a new visual maximum, we adopt a conservative value of 100 km s$^{-1}$, compatible with the wind velocity measured near the 1995 maximum. For a wind temperature of $10^4$ K and a distance of 6 kpc, we obtain a mass loss rate of $\dot M = (1.55\pm0.21)\times10^{-5}\, \mathrm{M}_{\sun}$ yr$^{-1}$. This mass-loss rate is in very good agreement with \cite{2009ApJ...698.1698G} estimates of the quiescent $\dot M$ of AG Car, of $1.5\times10^{-5}\,\mathrm{M}_{\sun}$ yr$^{-1}$.

\section{Conclusions}

We present APEX single-dish and ALMA/ACA interferometric observations of CO, $^{13}$CO and continuum towards AG Car, confirming the existence of a molecular structure associated with this LBV star. Below we summarize the key findings in this work:

\begin{enumerate}
    \item By means of CO and $^{13}$CO ALMA/ACA data, we report the detection of a molecular ring-like structure around AG Car, confirming the hypothesis by \cite{2002AJ....124.2920N}. The morpho-kinematic features of the structure can be explained by a slowly expanding ring or torus.
    \item We model the excitation conditions of the gas under non-LTE conditions. We used the available CO and $^{13}$CO $J=2\rightarrow1$ and CO $J=3\rightarrow2$ single-dish data to provide average estimates for the whole ring. The gas in the structure is warm, with a kinetic temperature of about 50 K, not very dense, with H$_2$ densities of a few $10^3$ cm$^{-3}$, and moderately thick. The total mass of molecular gas in the ring is 2.7 $\pm$ 0.9 $\mathrm{M}_{\sun}$, subject to uncertainties in the determination of the [CO/H$_2$] relative abundance. This accounts for at least a $\sim$30$\%$ of the mass expelled by the star.
    \item We develop a simple kinematic model to better study gas dynamics, showing that some deviations from pure radial expansion are required to properly reproduce some of the morpho-kinematic features observed in the data, such as: (1) the addition of a turbulent component or (2) the superposition of a differential rotational field.
    \item To explain the presence of a ring-like structure of molecular gas around AG Car, we discuss a number of possible formation scenarios, namely: (1) the compression of a remnant of the parent molecular cloud, (2) a RLOF mass transfer episode or a merger event in a close binary, or (3) an equatorial mass outflow. We regard the mass outflow as the most promising, taking into account the evolutionary stage of the star and the evidence available.
    \item We put together a global view of the mass-loss record of AG Car, proposing an overall morphology that integrates dust, ionised and neutral gas into a physically feasible, unified picture. We also derive a lower limit of the average mass-loss rate of the ring and compare it with the current mass-loss rate of the star derived from the ACA continuum.
\end{enumerate}

The detection of a molecular ring around AG Car adds the missing piece to the mass-loss history of this intriguing source. This work shows that molecules may trace a significant amount of the total mass lost by an evolved massive star. This has important implications for stellar evolution models, that sometimes struggle to properly reproduce the evolution between O/B and Wolf-Rayet stars, in part due to inaccurate mass-loss rate estimates.

Moreover, the results in this paper pave the way for further exploration of AG Car and its complex CSM. To confirm the origin of the ring, explain its survival and better understand its kinematics, observations at higher angular and velocity resolution are necessary. Looking for other molecules, such as PDR tracers, would also be extremely useful to obtain a more complete chemical characterization of the star. Finally, deeper radio continuum observations would provide valuable hints about the nebular excitation mechanisms and their dependence on the S-Dor cycles. AG Car is, undoubtedly, a unique object, and as such it represents an outstanding chance to learn about the nature of LBV phase and its associated mass-loss phenomena.

\section*{Acknowledgements}

We thank the anonymous referee for comments and suggestions that improved the quality of this paper. We are sincerely grateful to Dr. Damien Hutsemékers for kindly providing the H$\alpha$+[\ion{N}{ii}] and optical continuum EFOSC images of AG Car. This publication is based on data acquired with the Atacama Pathfinder Experiment (APEX). APEX is a collaboration between the Max-Planck-Institut fur Radioastronomie, the European Southern Observatory, and the Onsala Space Observatory. This paper makes use of the following ALMA data: ADS/JAO.ALMA$\#$2019.1.01056.S. ALMA is a partnership of ESO (representing its member states), NSF (USA) and NINS (Japan), together with NRC (Canada), MOST and ASIAA (Taiwan), and KASI (Republic of Korea), in cooperation with the Republic of Chile. The Joint ALMA Observatory is operated by ESO, AUI/NRAO and NAOJ. J.R.R. acknowledges support from projects ESP2017-86582-C4-1-R and  PID2019-105552RB-C4X (MINECO/MCIU/AEI/FEDER).

\section*{Data availability}

The derived data generated in this research will be shared on reasonable request to the corresponding author.

\bibliographystyle{mnras}
\bibliography{manuscript} % if your bibtex file is called example.bib

% Alternatively you could enter them by hand, like this:
% This method is tedious and prone to error if you have lots of references
%\begin{thebibliography}{99}
%\bibitem[\protect\citeauthoryear{Author}{2012}]{Author2012}
%Author A.~N., 2013, Journal of Improbable Astronomy, 1, 1
%\bibitem[\protect\citeauthoryear{Others}{2013}]{Others2013}
%Others S., 2012, Journal of Interesting Stuff, 17, 198
%\end{thebibliography}

%%%%%%%%%%%%%%%%%%%%%%%%%%%%%%%%%%%%%%%%%%%%%%%%%%

%%%%%%%%%%%%%%%%% APPENDICES %%%%%%%%%%%%%%%%%%%%%

%%%%%%%%%%%%%%%%%%%%%%%%%%%%%%%%%%%%%%%%%%%%%%%%%%

% Don't change these lines
\bsp	% typesetting comment
\label{lastpage}
\end{document}